\RequirePackage{lineno}
\documentclass[a4paper,12pt]{article}
\usepackage{slashed}
\usepackage[affil-sl,auth-sc]{authblk}
\usepackage{amssymb,amsmath,amsbsy}
\usepackage{bbm}
\usepackage{mathrsfs}
\usepackage{mathbbol}
\usepackage{bm}
\usepackage{appendix}
\usepackage{hyperref}
\usepackage[top=1.2in, bottom=1.1in,left=0.9in,includefoot]{geometry}
\usepackage{subcaption}
\usepackage{graphicx}
\usepackage{cite}
\usepackage{float}
\usepackage{caption}
\usepackage{wasysym}
\usepackage{amsthm}
\usepackage[utf8x]{inputenc}
\usepackage[english]{babel}
\usepackage[capitalize]{cleveref}
\newif\ifContLineTwo
\newif\ifContLineThree
\def\conC#1{\vbox{\ialign{##\crcr
  \ifContLineThree\hrulefill\else\vphantom{\hrulefill}\fi\crcr
  \noalign{\kern3.2pt\nointerlineskip}
  \ifContLineTwo\hrulefill\else\vphantom{\hrulefill}\fi\crcr
  \noalign{\kern3.2pt\nointerlineskip}
  \ifContLineOne\hrulefill\else\vphantom{\hrulefill}\fi\crcr
  \noalign{\nointerlineskip}
  $\hfil\textstyle{\vbox to 14pt{}#1}\hfil$\crcr}}}
\def\DrawLeg#1#2{
  \kern-.2pt              
  \dimen2 =#1             
  \advance\dimen2 by 2pt  
  \dimen3 = 10.6pt        
  \dimen4 =3.6pt          
  \advance\dimen3 by -\dimen2 
  \multiply\dimen4 by #2
  \advance\dimen3 by \dimen4
  \raise\dimen2 \hbox{\vrule height\dimen3 width .4pt} 
  \kern-.2pt}             
\def\begC#1#2{\setbox0 =\hbox{$\textstyle{#2}$}
  \dimen0=.5\wd0 \dimen1=\ht0
  \conC{\hskip\dimen0}
  \count255=#1
  \ifnum\count255 =1 \ContLineOnetrue\else
  \ifnum\count255 =2 \ContLineTwotrue\else
  \ifnum\count255 =3 \ContLineThreetrue\fi\fi\fi
  \DrawLeg{\dimen1}{\count255}
  \conC{\hskip\dimen0}
  \kern-\dimen0\kern-\dimen0 \box0}
\def\endC#1#2{\setbox0 =\hbox{$\textstyle{#2}$}
  \dimen0=.5\wd0 \dimen1=\ht0
  \conC{\hskip\dimen0}
  \count255=#1
  \ifnum\count255 =1 \ContLineOnefalse\else
  \ifnum\count255 =2 \ContLineTwofalse\else
  \ifnum\count255 =3 \ContLineThreefalse\fi\fi\fi
  \DrawLeg{\dimen1}{\count255}
  \conC{\hskip\dimen0}
  \kern-\dimen0\kern-\dimen0 \box0}

\def\lsim{\buildrel{\scriptscriptstyle <}\over{\scriptscriptstyle\sim}}
\def\gsim{\buildrel{\scriptscriptstyle >}\over{\scriptscriptstyle\sim}}

\newcommand{\bea}{\begin{eqnarray}}
\newcommand{\eea}{\end{eqnarray}}

\usepackage{color}

\definecolor{orange}{rgb}{0.9,0.2,0}
\definecolor{brown}{rgb}{0.7,0.3,0.2}
\definecolor{fuxia}{rgb}{1,0,1}
\definecolor{skyblue}{rgb}{0,0.1,0.9}
\definecolor{violetred}{rgb}{0.8,0.13,0.56}
\definecolor{deeppink}{rgb}{1.00,0.08,0.5}
\definecolor{pink}{rgb}{1.00,0.75,0.80}
\definecolor{orchid}{rgb}{0.85,0.44,0.84}
\definecolor{lightpink}{rgb}{1.00,0.71,0.76}
\definecolor{bluish}{rgb}{0,0.6,0.8}
\numberwithin{equation}{section}
\title{\bf Light Singlino Dark Matter at the LHC}
\author{\small Monoranjan Guchait \thanks{guchait@tifr.res.in}~}
\author{\small Arnab Roy \thanks{arnab.roy@tifr.res.in}~}
\affil{\small Department of High Energy Physics, \\
Tata Institute of Fundamental Research, \\
 Homi Bhabha Road, Mumbai-400005, India}

\def \MET{E{\!\!\!/}_T}
\def\invfb{\text{fb}^{-1}}

\def\mh{\rm{m_{H_1}}}
\def\ma{\rm{m_{A_1}}}
\def\mhh{\rm{m_{H_2}}}
\def\maa{\rm{m_{A_2}}}
\def\mhhh{\rm{m_{H_3}}}
\def\muf{\rm{\mu_{eff}}}
\def\tb{$\rm {tan\beta}$~}
\def\hu{{$\rm {H_u}$}}
\def\hd{$\rm{H_d}$}

\def\t1 {\widetilde {t_1}}
\def\N1{\widetilde \chi_1^0}
\def\N2{\widetilde \chi_2^0}
\def\N3{\widetilde \chi_3^0}
\def\N0{\widetilde \chi^0}
\def\C1{\widetilde \chi_1^{\pm}}
\def\mst1 {m_{\t1}}
\def\br {\begin{eqnarray}}
\def\er {\end{eqnarray}}

\def\lumi{\cal L}
\def\lsim{\buildrel{\scriptscriptstyle <}\over{\scriptscriptstyle\sim}}
\def\gsim{\buildrel{\scriptscriptstyle >}\over{\scriptscriptstyle\sim}}

\date{}

\begin{document}
\maketitle
\begin{abstract}
The light singlino-like neutralino is expected to be a promising candidate for DM in the allowed parameter space
of the NMSSM. The DM annihilation process takes place via 
the light Higgs bosons which are natural in this model.  
Identifying the allowed region of parameter space 
including various constraints,  
the detection prospect of such light DM candidate and
Higgs bosons are explored at the LHC with its high 
luminosity options. Light Higgs bosons and the DM candidate, the lightest 
singlino-like neutralino are 
indirectly produced at the LHC via the SM Higgs production 
and its subsequent decays.
Jet substructure techniques are used to tag boosted Higgs. It is found that the favourable range of masses of Higgs bosons and neutralino, compatible with a low mass DM solution, can be discovered with a reasonable signal significance ($\sim 5 \sigma$) at the LHC, with the center of mass energy $\sqrt{s}=14$ TeV and integrated luminosity options ${\cal L}=$300~$\invfb$ and 3000~$\invfb$.
\end{abstract}
\vskip .5 true cm
\newpage

\section{Introduction}
Understanding the nature of dark matter(DM) candidate is of great interest in the present day of particle physics, particularly, in the context of beyond standard model(BSM) physics. Huge efforts are in place by various experiments to search for DM candidate via direct and indirect manner\cite{Hooper:2008sn,Feng:2010gw}. Unfortunately, still the candidate of DM remains elusive. Very recent observations from the PLANCK~\cite{Aghanim:2018eyx} experiment predict the limits of relic density 
at 2$\sigma$ as, 
\br
\rm \Omega h^2=0.12~\pm~0.001.
\label{eq:wmap}
\er 
It is observed that the DM annihilation cross-section at the weak scale naturally predicts relic density consistent with this PLANCK data. Currently, searches for DM candidates are one of the most exciting and challenging programs. Numerous dedicated experiments including the Large Hadron Collider (LHC) are involved in this endeavour, and have made considerable progress. However, all negative results in direct searches of DM experiments, lead to stringent limits on DM-nucleon 
scattering cross-sections in terms of DM particle masses\cite{Aprile:2018dbl,Akerib:2016vxi,Cui:2017nnn,Agnes:2018ves,Aprile:2019dbj,Amole:2019fdf}.
As we know, because of the non-relativistic nature of DM candidate, 
the DM-nucleon scattering cross-section can be separated into two parts, spin-independent(SI) and spin-dependent(SD). 
The SI part is mediated by scalars and increases 
with the mass of nucleon, whereas the SD process 
involving axial-vector coupling with nuclear spin is mediated by 
gauge bosons. Obviously, the SD cross-section is larger than SI 
because of the suppressed coupling due to light 
quark masses~\cite{Barger:2008qd,Belanger:2008sj,Agrawal:2010fh}.  
Recent measurements by the XENON1T experiment reported an upper limit 
of the DM-nucleon SI elastic scattering cross-section at 
4.1$\times 10^{-47}{\rm cm}^2$ and  
2$\times 10^{-44}{\rm cm}^2$ corresponding to DM particle masses 
of 30 GeV and 6~GeV respectively~\cite{Aprile:2018dbl}. 
These are the most stringent limits to date, whereas limits 
from LUX\cite{Akerib:2016vxi} and PANDA\cite{Cui:2017nnn} 
are not competitive. 
With the detector upgrade in XENONnT experiment, the sensitivity
is expected to improve by an order of magnitude~\cite{Aprile:2018dbl}.
It is to be noted that the XENON1T experiment
is not sensitive to the lower range($<$6~GeV) of DM particle masses.
However, there are few other experiments that are sensitive 
to this lower mass range of DM~\cite{Aalseth:2010vx,Agnese:2013rvf,Bernabei:2013xsa,Agnes:2018ves}. For instance, DarkSide-50 experiment   
searches for DM candidate covering the mass 
range $\lsim20$ GeV, and lack of observation of any signal 
event leads to an exclusion limit on DM-nucleon SI cross-section 
at $10^{-41}{\rm cm}^2$, at 90\% C.L, corresponding to the DM particle of mass 
1.8 GeV~\cite{Agnes:2018ves}. 
Similarly, the SD DM-proton and DM-neutron scattering cross-sections 
are also constrained for a reasonably wide range of DM particle masses. The exclusion 
limit on SD DM-neutron scattering cross-section also come 
from XENON1T, which predicts at 90\% CL, an upper limit 
6.3$\times 10^{-42}{\rm cm}^2$ for 30 GeV DM particle mass, and 
it increases further to  3$\times 10^{-39}{\rm cm}^2$ for 
6 GeV mass~\cite{Aprile:2019dbj}. 
The most stringent SD cross-section limit to date on DM-proton scattering
cross-section at 90\% C.L is $\sim 3\times 10^{-41}{\rm cm}^2$ for 
20 GeV DM particle mass which comes from the PICO-60~\cite{Amole:2019fdf} experiment. 
Apart from these direct searches,  DM candidates are also explored 
indirectly at the LHC experiment. The DM particle produced in 
proton-proton collision at the LHC leaves an imbalance of missing 
energy signature in the detector because of its extremely 
weak interaction with matter. 
Hence, the final state consisting of a hard missing energy along with 
a recoil of visible energy is assumed to be a classical 
signature of DM.    
Currently, in both CMS and ATLAS experiments, searching for 
the signature of DM candidates are treated as a 
high priority analysis~\cite{Abercrombie:2015wmb}. 
However, from the non-observation of any signal events in data, 
model-dependent limits of DM particle masses are set by both CMS and 
ATLAS experiments~\cite{ATLAS:2018iyk,Aaboud:2019yqu}. Evidently, even in the 
presence of stringent constraints on DM particle masses from all 
direct and indirect searches, as discussed above, still a 
considerable range of lower ($\sim$ few GeV) and 
higher (${\cal O}(100)$~GeV) range of masses are not ruled out. 
Naturally, this observation attracts special attention to 
look for models, which can offer viable 
DM candidates of those mass ranges compatible with data. 
Motivated by this fact, in this paper
we try to find models of DM particle corresponding to this
lower range of masses, 
which can provide solution consistent with all 
constraints due to direct and indirect searches, as discussed 
above\cite{Aprile:2018dbl,Akerib:2016vxi,Cui:2017nnn,Agnes:2018ves,Aprile:2019dbj,Amole:2019fdf,ATLAS:2018iyk,Aaboud:2019yqu}.

Variety of well-motivated BSM predicts a plethora of cold and warm DM particle 
candidates~\cite{Bertone:2004pz,Belanger:2009br}. Among them, the most 
widely studied DM model is offered by minimal supersymmetric 
standard model (MSSM), where the most popular candidate for DM with 
conserved R-parity is the lightest neutralino ($\N0_1$), a majorana 
spin 1/2 particle. In order to obtain right relic 
density~(Eq.\ref{eq:wmap}), the DM candidate is favoured to be 
the Higgsino-like, and of the mass range $\sim$1  
TeV~\cite{ArkaniHamed:2006mb,Baer:2016ucr,Chakraborti:2017dpu}. Interestingly, assuming neutralino as a thermal relic, the relic density bound sets a lower limit on the neutralino mass, $\rm m_{\tilde{\chi}_1^0}\gsim 34~GeV$, in the framework of MSSM~\cite{Barman:2017swy}. Naturally, it rules out the possibility of having any DM candidate of very low mass ($\sim$ few GeV) in MSSM. Like the MSSM, the theory of next-to-minimal supersymmetric 
standard model(NMSSM)~\cite{Fayet:1974pd,Drees:1988fc,Ellis:1988er,Ellwanger:2009dp} offers the lightest neutralino as a 
potential DM candidate. 
In the NMSSM, the Higgsino mass term($\mu$) is generated dynamically, 
in order to cure the $\mu$-problem~\cite{KIM1984150}, by adding a 
singlet scalar field with two Higgs doublet and extending the Higgs sector 
resulting in seven Higgs bosons states. 
Because of the interplay between model parameters in the Higgs sector, one or two of the 
Higgs boson states can be very light, even less than the mass of the
SM-like Higgs boson, without violating any collider 
constraints\cite{Ellwanger:2011sk,Christensen:2013dra,Cao:2013gba,Guchait:2015owa,Domingo:2015eea,Kumar:2016vhm,Guchait_2017}. 
Furthermore, the singlino, fermionic superpartner of singlet 
field, extends the neutralino sector with five physical states, where the 
lightest neutralino state plays the role as a DM candidate.  
In particular, even with a very low mass ($\sim$ few GeV), 
the neutralino, favourably to be singlino-like, appears as a viable dark matter 
candidate without violating any existing constraints predicted by several
DM experiments\cite{Belanger:2005kh,Gunion:2005rw,Hugonie:2007vd,Ellwanger:2014hia,Ellwanger:2016sur,Mou:2017sjf,Ellwanger:2018zxt,Abdallah:2019znp,Baum:2017enm}.
In such a scenario, the possible DM annihilation process 
occurs via light Higgs bosons reproducing right cross-section 
consistent with the relic density given by Eq.~\ref{eq:wmap}. 
This phenomena resembles the scenario of Higgs portal model, 
where light Higgs boson acts as a portal between the SM and non-SM 
sector~\cite{Arcadi:2019lka}.
Moreover, the DM(singlino)-nucleon scattering cross-sections, both SI and SD, 
satisfy experimental limits predicted by direct searches, 
thanks to the presence of appropriate singlino composition in the 
lightest neutralino state. In this regard, naturally, the immediate and 
pertinent question to ask is about the prospect of detecting 
the signal of this low mass singlino-like DM candidate 
at the LHC. In literature, quite a few studies exist in this context predicting the discovery potential of DM candidate at the LHC\cite{Huang:2013ima,Huang:2014cla,Potter:2015wsa,Xiang:2016ndq,Ellwanger:2018zxt,Abdallah:2019znp}. The objective of this present study is to revisit this DM solution 
in the framework of NMSSM, and then explore the detection prospect 
of such a scenario at the LHC for the current and future luminosity options. 
More precisely, our goal is to find discovery potential of 
light singlino-like neutralino and Higgs boson states at the LHC, 
which in combination provide a right DM solution.

At the LHC, the direct production 
of light singlino state and singlet-like Higgs bosons, 
having negligible coupling with fermions and gauge bosons 
are very much suppressed. In such a scenario, 
these particles can be produced indirectly via the production of 
some other intermediate particles which are having non-negligible couplings with those 
states~\cite{Guchait_2016,Guchait_2017,Guchait:2017ztk,Abdallah:2019znp}.  
For example, in this paper, we consider the production of non-SM-like 
light Higgs bosons via the decays of SM Higgs boson 
which is produced through a standard mechanism.
Subsequently, light singlino states are  produced via the 
decay of light non-SM-like Higgs bosons. 
It is to be noted that the corresponding branching ratios(BR) 
of all these decay modes are very much sensitive to model 
parameters, which will be discussed in detail in later sections.

The SM Higgs boson is considered to be produced via gluon-gluon fusion, 
which is the dominant production 
mechanism~\cite{Heinemeyer:2013tqa,Dittmaier:2011ti}. 
In order to give a boost to the final state, the SM Higgs boson is 
produced exclusively along with a jet. 
Consequently, the pair of lighter Higgs boson states originating from the decay of SM Higgs boson of mass 125 GeV are moderately boosted ($\rm p_T \sim 30-40$ GeV) with a reasonable separation between them, and so the decay products from those states emerge as a collimated object. Thus the jets coming from light Higgs decay appear as a single fat jet. Using jet substructure technique this ``Higgs jet'' (HJ) is tagged where the two subjets are likely to be b-like~\cite{Butterworth:2008iy,Asquith:2018igt}. 
In summary, we focus on the signal final state consisting 
an HJ and missing energy, along with at least 
one untagged QCD jet. Considering this signal final state, 
we perform detail signal and SM background simulation 
and predict signal sensitivity for 300~$\invfb$ and 
3000~$\invfb$ luminosity options.

The paper is organised as follows. In section 2, we review the NMSSM model briefly and discussed the region of parameter space compatible with relic density and DM-nucleon scattering constraints as well. The relevant range of parameters are identified through a numerical scan. Signal and background simulation is presented in section 3, followed by a discussion on the results. Finally, the summary is presented in section 5.

\section{The NMSSM Model and Dark Matter relic density}
\label{sec:model}
In this section, we briefly outline the NMSSM model set up relevant to our scenario, which provides light singlet-like Higgs bosons and a light singlino-like neutralino as a DM candidate with right relic density (Eq.\ref{eq:wmap}). The NMSSM contains an additional gauge singlet superfield (S) along with two Higgs doublet superfields (\hu~and \hd). The corresponding $Z_3$-invariant superpotential is given by \cite{Fayet:1974pd,Drees:1988fc,Ellis:1988er,Ellwanger:2009dp,Miller:2003ay}, 
\br
\rm W_{NMSSM}= W_{MSSM}+ {\lambda S H_u H_d +\frac{1}{3}\kappa {S}^3},
\er        
where $\lambda$ and $\kappa$ are the dimensionless couplings, and $\rm{W_{MSSM}}$ represents the part of the superpotential in MSSM counting Higgs doublets but without the $\mu$-term. 
In addition, two soft terms, $\rm{\lambda A_{\lambda}SH_u H_d}$ and $\rm{\frac{1}{3}\kappa A_{\kappa} S^3}$ are also included. 
The Yukawa like term with coupling $\lambda$ generates the Higgsino mass term, $\muf=\lambda$$v_s$, where $v_s$ is the vacuum expectation value (VEV) acquired by the singlet superfield. The dynamic generation of the $\mu$-term, the key aspect for the motivation of NMSSM, prevents it to acquire large value \cite{KIM1984150}. The Higgsino mass term is expected to be at the level of the electroweak scale to obtain appropriate electroweak symmetry breaking~\cite{Ellis:1988er}. On the other hand, phenomenologically, $\muf$ is restricted to be $\muf\gsim100$ GeV, due to the chargino mass limit predicted by LEP experiment \cite{LEPSUSY}.

As mentioned before, the enlarged Higgs sector of NMSSM consists of seven physical Higgs bosons, 3 CP even states ($\rm H_1, H_2, H_3,$ assuming $\mh < \mhh < \mhhh$) and 2 CP odd states ($\rm A_1, A_2,$ assuming $\ma < \maa$) and 2 charged Higgs($\rm{H^{\pm}}$) boson states. Masses and couplings of these Higgs bosons are determined by model parameters. The Higgs sector is briefly revisited here to identify respective ranges of corresponding parameters to our interest. The 3 CP even Higgs states are described by 3$\times$3 mass matrices in the basis $\rm{\psi_R \equiv (H_{uR}, H_{dR},S_R)}$, the real parts of Higgs fields. The elements of mass matrix are given by \cite{Ellwanger:2009dp},
\begin{eqnarray}
\rm M_{S,11}^2 &=& \rm M_Z^2\sin^2\beta + \muf\cot\beta ( A_{\lambda} + \kappa v_s ),\nonumber\\
\rm M_{S,22}^2 &=& \rm M_Z^2\cos^2\beta + \mu_{eff}\tan\beta(A_{\lambda} + \kappa v_s),\nonumber\\
\rm M_{S,33}^2 &=& \rm \frac{\lambda^2v^2 A_\lambda\sin2\beta}{2\mu_{eff}}
+ \kappa v_s(A_{\kappa}+4\kappa v_s),
\label{eq:CPevenHiggs}\\
\rm M_{S,12}^2 &=& \rm(\lambda^2 v^2 - \frac{M_Z^2}{2} )\sin2\beta - \mu_{eff}(A_\lambda + \kappa v_s), \nonumber\\
\rm M_{S,13}^2 &=& \rm\lambda v (2\mu_{eff} \sin\beta - (A_\lambda+2 \kappa v_s)\cos\beta),\nonumber\\
\rm M_{S,23}^2 &=& \rm\lambda v (2\mu_{eff} \cos\beta - (A_\lambda+2 \kappa v_s)\sin\beta).\nonumber
\end{eqnarray}
Here \tb is the ratio of VEVs of neutral components of two Higgs doublet. The masses of 3 CP-even Higgs boson states can be obtained by diagonalising the mass matrix by an orthogonal matrix ($\rm S_{ij}$; i,j=1-3), and hence physical states ($\rm{H_i}$) become the admixture of weak Higgs boson states as,
\br
\rm{H_i = \sum_{j=1}^3 S_{ij} \psi_{jR}}.
\er     
Notably, the lightest CP-even Higgs boson mass is found to be bounded by \cite{MOROI199273,Franke:1995tc}, $\rm{m_{H_1}^2 \le M_Z^2\cos^22\beta + \lambda^2} $$v_s^2$$\rm \sin^22\beta$ at the tree level. Notice that the extra contribution lifts the tree level Higgs boson mass substantially, and hence may not require a huge contribution from higher-order correction \cite{King:2012is}. As a consequence, a wide region of parameter space which is less constrained can easily accommodate one of the CP-even Higgs boson (primarily either $\rm{H_1}$ or $\rm{H_2}$) states as the SM-like Higgs boson with a mass $\sim$ 125 GeV. This feature makes the NMSSM very attractive after the discovery of the SM Higgs boson at the LHC~\cite{King:2012is,Agashe:2012zq,Vasquez:2012hn,Badziak:2013bda,Guchait:2015owa}.

In the CP-odd sector, eliminating the Goldstone modes, the elements of 2$\times$2 mass matrix for CP odd Higgs boson states in $\rm{\psi_I \equiv (A,S_I)}$ basis are given as,  
\begin{eqnarray}
\rm M^2_{P,11}&=& \frac{2\muf}{\sin2\beta} (A_{\lambda} + \kappa v_s),\nonumber\\
\rm M^2_{P,22}&=&\lambda^2 v^2 \frac{\sin2\beta}{2\muf}(A_{\lambda}+4\kappa v_s) 
- 3 A_\kappa\kappa v_s, 
\label{eq:CPodd}\\
\rm M^2_{P,12}&=&\lambda v(A_\lambda -2 \kappa v_s). \nonumber
\end{eqnarray}
Similarly, diagonalising this mass matrix by an orthogonal ($\rm{P_{ij}}$, i,j = 1,2) matrix, the masses of the two physical CP-odd states ($\rm{A_1, A_2}$) can be obtained, and hence the corresponding composition of physical states are given as,
\br
\rm{A_i =\sum_{j=1}^2 P_{ij}\psi_{jI}}.
\label{cpodd}
\er   
Interestingly, unlike the MSSM, in NMSSM, the physical Higgs boson states contain a fraction of the singlet component ($\rm S_I$) which does not couple with fermions and gauge bosons. Of course, the content of singlet component in physical states is very much parameter space sensitive. 

The Higgs sector and the corresponding masses and composition of physical states are described by six parameters:
\begin{equation}
\rm{\lambda, \kappa, A_{\lambda}, A_{\kappa}, \tan\beta, \mu_{eff}}.
\label{eq:param}
\end{equation}
Dependence on squark masses and other trilinear terms ( A-terms) occurs via radiative corrections\cite{Hall:2011aa}.

The fermionic superpartner ($\rm \tilde{S}$) of the singlet field, mixes with Higgsinos extending the neutralino mass matrix to $5 \times 5$, in the basis ($-i\rm{\tilde B}$,$ -i\rm{\tilde W_3}$,$\rm \tilde{H}_u^0, \tilde{H}_d^0, \tilde{S}$) and it is presented as,
\br
\rm M_N = \left( \begin{array}{ccccc}
	\rm M_1 &0 &\frac{-g_1 v s_{\beta}}{\sqrt 2} & \frac{g_1 v c_{\beta}}{\sqrt 2}  & 0  \\
	0 &\rm M_2   & \frac{g_2 v s_{\beta}}{\sqrt 2} & \frac{-g_2 v c_{\beta}}{\sqrt 2} &0 \\
	\frac{-g_1 v s_{\beta}}{\sqrt 2} & \frac{g_2 v s_{\beta}}{\sqrt 2}  & 0 & - \muf &  -\lambda v c_{\beta}\\
	\frac{g_1 v c_{\beta}}{\sqrt 2}   & \frac{-g_2 v c_{\beta}}{\sqrt 2}  & - \muf  & 0 &  - \lambda v s_{\beta} \\
	0 & 0 &  -\lambda v s_{\beta} &  - \lambda v c_{\beta}  &  2 \kappa v_s \end{array} \right),
\label{eq:mneu}
\er
with $\rm s_{\beta}\equiv sin \beta,\; c_{\beta}\equiv cos\beta$, $\rm M_1$ and $\rm M_2$ are the masses of $\rm \tilde B$ and $\rm \tilde W_3$ gauginos respectively, $v_u$ and $v_d$ are the VEVs for neutral components of $\rm H_u$ and $\rm H_d$ fields and are constrained to be $v_u^2 + v_d^2=v^2$; $\rm g_1$ and $\rm g_2$ are weak couplings. The masses of 5 neutralino states, $\rm m_{\tilde{\chi}_i^0} (i=1,..,5)$ can be obtained by diagonalising the mass matrix $\rm M_N$ by an orthogonal matrix $\rm N_{5\times 5}$ as,
\br
\rm M_{\tilde \chi^0}^D  = \rm N M_N N^{\dagger}.
\label{eq:nn}
\er      
The analytical expressions of $\rm m_{\tilde{\chi}^0_i}$ and the corresponding physical states  exist in the literature for the MSSM\cite{Guchait:1991ia,Choi:2001ww}, and as well as for the NMSSM \cite{Pandita:1994ms,Choi:2004zx}. The masses and couplings of neutralinos are very sensitive to NMSSM specific parameters, in particular $\lambda$, $\kappa$ and $v_s$ or $\rm \mu_{eff}$, along with $\rm M_1$ and $\rm M_2$. Moreover, these parameters (except $\rm M_1$ and $\rm M_2$) are also strongly connected with the Higgs sector
(Eq.~\ref{eq:CPevenHiggs} -- \ref{eq:CPodd}), and play important roles, 
along with $\rm A_\lambda$ and $\rm A_\kappa$ in determining 
the masses and mixings of Higgs bosons.

As stated earlier, the goal of this study is to provide a low 
mass DM solution within the framework of the NMSSM. With this motivation,
we try to identify the corresponding regions of relevant model 
parameters compatible with all existing experimental constraints.

In our proposed solution, DM annihilation takes place via s-channel 
mediated by light Higgs scalars giving a pair of fermions in the final state~\cite{Kozaczuk:2013spa,Han:2014nba,Wang:2020dtb},
\br
\rm \chi\chi \to{H_1/A_1} \to f \bar f.
\label{eq:anni}
\er
The DM annihilation rate is primarily sensitive to the interaction  
between neutralino pair and Higgs boson, and their relative mass difference. 
The Higgs-neutralino-neutralino couplings 
are given as\cite{Ellwanger:2009dp,Gunion:2005rw},
\begin{equation}
\begin{aligned}
\rm g_{{\N0_1}{\N0_1} H_i}= \sqrt{2}\lambda N_{15}(S_{i1}N_{14}+S_{i2}N_{13})+\sqrt{2}S_{i3}(\lambda N_{13} N_{14} - \kappa N_{15}^2 ) \\ \rm -\frac{g}{2}(N_{12} - \tan\theta_w N_{11})(S_{i1} N_{13} - S_{i2}N_{14}) ,
\end{aligned}
\label{eq:ghnn}
\end{equation}

\begin{equation}
\begin{aligned}
\rm g_{{\N0_1}{\N0_1} A_i}= \sqrt{2}\lambda N_{15}(P_{i1}N_{14}+P_{i2}N_{13}) 
+\sqrt{2}P_{i2}(\lambda N_{13} N_{14} - \kappa N_{15}^2 ) \\
\rm -\frac{g}{2}(N_{12} - \tan\theta_w N_{11})(P_{i1} N_{13} - P_{i2}N_{14}) ,
\label{eq:gann}
\end{aligned}
\end{equation}  
Here $\rm N_{15}$ presents the singlino composition of the lightest neutralino, 
whereas $\rm S_{i3}$ and $\rm P_{i2}$ stand for the singlet 
components of $\rm H_i$ and $\rm A_i$ respectively. Parameters 
$\lambda$ and $\kappa$, which are connected with the singlino mass 
and its composition, are found to be very sensitive to the  
annihilation cross-section due to the above couplings(Eq.~\ref{eq:ghnn} and 
\ref{eq:gann}). The analytical expressions for the cross-section of annihilation 
processes are presented in Appendix A. As indicated, the right relic 
density corresponding to the lower range ($\lsim$ 20 GeV) of DM masses 
can be achieved by requiring neutralino and Higgs boson states 
singlino and singlet dominated 
respectively (i.e. $\rm N_{15}$,$\rm P_{12}, S_{13}\sim$ 1), for which
\br
\rm g_{{\N0_1}{\N0_1} H_1} \sim \rm \sqrt{2} S_{13} (\lambda N_{13}N_{14} - \kappa N_{15}^2)
\rm \sim \rm - \sqrt{2} S_{13} \kappa N_{15}^2,
\label{eq:ghnn1}
\er

\begin{equation}
\rm g_{{\N0_1}{\N0_1} A_1} \sim - \sqrt{2} P_{12} \kappa N_{15}^2.
\label{eq:gann1}
\end{equation}
The DM-nucleon scattering cross-sections, both $\rm \sigma_{SI}$ 
and $\rm \sigma_{SD}$ mediated by Higgs scalars and gauge bosons 
respectively, are given in Appendix B. 
From direct searches, allowed spin-independent cross-section 
corresponding to DM masses of our interest, varies 
from $\rm \sim 10^{-44}\; cm^2 - 10^{-46} \; cm^2$, 
which is achievable through the adjustments of 
coupling $\rm g_{{\N0_1}{\N0_1} H_1}$ or $\rm g_{{\N0_1}{\N0_1} A_1}$. 
Again, we observed that a singlino-like lightest neutralino and singlet dominant light Higgs bosons are most favoured. It suggests that the light singlino-like DM candidate requires singlet dominated light Higgs boson states in order to have right relic density and DM-nucleon scattering cross-section \cite{Aghanim:2018eyx,Aprile:2018dbl,Agnes:2018ves,Aprile:2019dbj,Amole:2019fdf}. Therefore, the preferred parameter space favouring our scenario should provide, 
(a) a light singlino-like LSP, (b) light singlet-like Higgs boson states.

A closer look at the neutralino mass matrix reveals few features of neutralino masses and mixings~\cite{Miller:2003ay}. For instance, the absence of mixing terms between singlino and gaugino fields implies no interaction between singlino-like neutralino and 
gaugino-like or gauge boson states. Notice that the mixing 
between singlet and doublet Higgs fields is decided by 
$\lambda v\rm \cos\beta$ or $\lambda v\rm \sin\beta$ (Eq.~\ref{eq:mneu}). 
Among the five neutralino states, two of them remain to be gaugino-like if, $\rm |M_{1,2}-\mu_{eff}|\geq M_Z$, the mass of Z-boson. 
For a decoupling scenario, $ 2|\kappa| v_s << \rm \mu_{eff}, M_{1,2}$, 
the mass of singlino-like neutralino turns out to 
be $\sim 2 |\kappa| v_s$, and dominantly a singlino-like. 
On the other hand, since $\muf$ or $\lambda v_s\rm \sim {\cal O}(100)$~GeV, 
hence for smaller values of $\lambda \lsim$0.1, the typical value 
of $v_s$ is expected to be large $\sim{\cal O}(1)$~TeV. Therefore, 
for a very light singlino-like LSP, $|\kappa|$ should lie within the  
range of $\sim10^{-3}$. For higher values of $\lambda \sim 0.1 $, it is 
possible to accommodate comparatively lower values of $v_s$, with 
little larger values of $|\kappa|$ \footnote{Requirement of both $\kappa$ and $\lambda$ remain to be perturbative up to GUT scale impose the constraint $\lambda^2 + \kappa^2 \lsim$0.5\cite{Miller:2003ay}.}. In fact, the mass of singlino-like 
LSP, $\rm m_{\N0_1}$ $\sim$ 2$\left(\frac{\kappa}{\lambda}\right){\muf}$ 
becomes small for $\frac{\kappa}{\lambda} \sim 10^{-2}$. On the contrary, 
for  $\rm 2 |\kappa| v_s >> M_{1,2},\mu_{eff}$, singlino-like state 
becomes very heavy, and decouples from other neutralino states which 
consist of only Higgsino and gaugino components like MSSM scenario. 
The other NMSSM parameters $\rm A_\kappa$ and $\rm A_\lambda$, which are 
not related with neutralino masses and mixings at the tree level, 
are expected to be restricted due to the requirements of light 
singlet-like Higgs bosons. Following Eq.~\ref{eq:CPodd}, the lighter CP odd state($\rm A_1$) is found to be singlet-like for decoupling type of scenario such as \cite{Guchait:2016pes},
\br
\rm M^2_{P,11} > M^2_{P,12}, M^2_{P,22}, 
\er
which also leads heavier state ($\rm A_2$) MSSM like.\footnote{Similar scenario can also occur if off-diagonal entry $\rm M^2_{P,12} \sim 0$. But it is 
	not a viable option to our interest as $ \kappa v_s$ is required 
	to be very small.} Hence, $\rm A_{\lambda}$ is preferred to be very 
large ($\sim$ 2-3 TeV) (see Eq.~\ref{eq:CPodd}), since 2$|\kappa|v_s$ 
is required to be very small, to have a light LSP. 
With a good approximation, one can obtain mass of $\rm A_1$ as 
$\rm m_{A_1}^2 \simeq -3 A_\kappa$$\kappa v_s$ \cite{Miller:2003ay}. Moreover, as required above, $|\kappa| v_s$ cannot be large, so a moderate range (${\cal O} $(10) GeV) of $\rm A_\kappa$ is required to obtain a light $\rm A_1$ state. For the CP even Higgs sector, the spectrum of relevant parameters corresponding to our interest can be understood following a sum rule obtained using the tree level masses of $\rm H_1$ and $\rm H_2$. This sum rule reads as \cite{Miller:2003ay},
\br
\rm m_{H_1}^2 +m_{H_2}^2 \equiv M_Z^2 + \frac{1}{2} \kappa v_s
(4 \kappa v_s +\sqrt{2}A_\kappa).
\label{eq:sumrule}
\er         
Naturally, if any of the Higgs boson states (here it is $\rm H_2$) becomes massive and close to the mass of the SM-like Higgs boson, then for a moderate value of $\rm A_{\kappa}$, $\rm H_1$  state becomes very light, even may be less than the half of the mass of $\rm H_2$ state. Since $\rm H_2$ state is SM-like, hence mixing between singlet and doublet components ($\rm H_u,H_d$) should be very small, yielding  $\rm H_1$ state mostly singlet dominated.  
Furthermore, since the annihilation process occurs via s-channel 
Higgs exchange, the cross-section enhanced significantly, 
for $\rm m_{A_1/H_1} \sim 2\times m_{\N0_1}$, which we also 
require for our proposed collider searches. The third CP even physical Higgs 
state $\rm H_3$, seems to be very massive and decoupled for 
large values of $\rm A_\lambda$. 
Finally, with all these above arguments corresponding to our proposed 
scenario, we conclude :

\begin{itemize}
	\item
	light singlino-like LSP requires  very small $ |\kappa| v_s$, with $\rm \kappa/\lambda \sim 10^{-2}$,
	\item
	requirement of light Higgs boson states to be singlet-like, leads  $\rm A_\lambda$ to be very large(few TeV),
	but $\rm A_\kappa$ not necessarily to be very large, but with a relative sign opposite to $\kappa$.  
\end{itemize}

\section{Parameter scan}
\label{sec:scan}
\begin{figure}[H]                  
	\centering
	\includegraphics[width=0.5\linewidth]{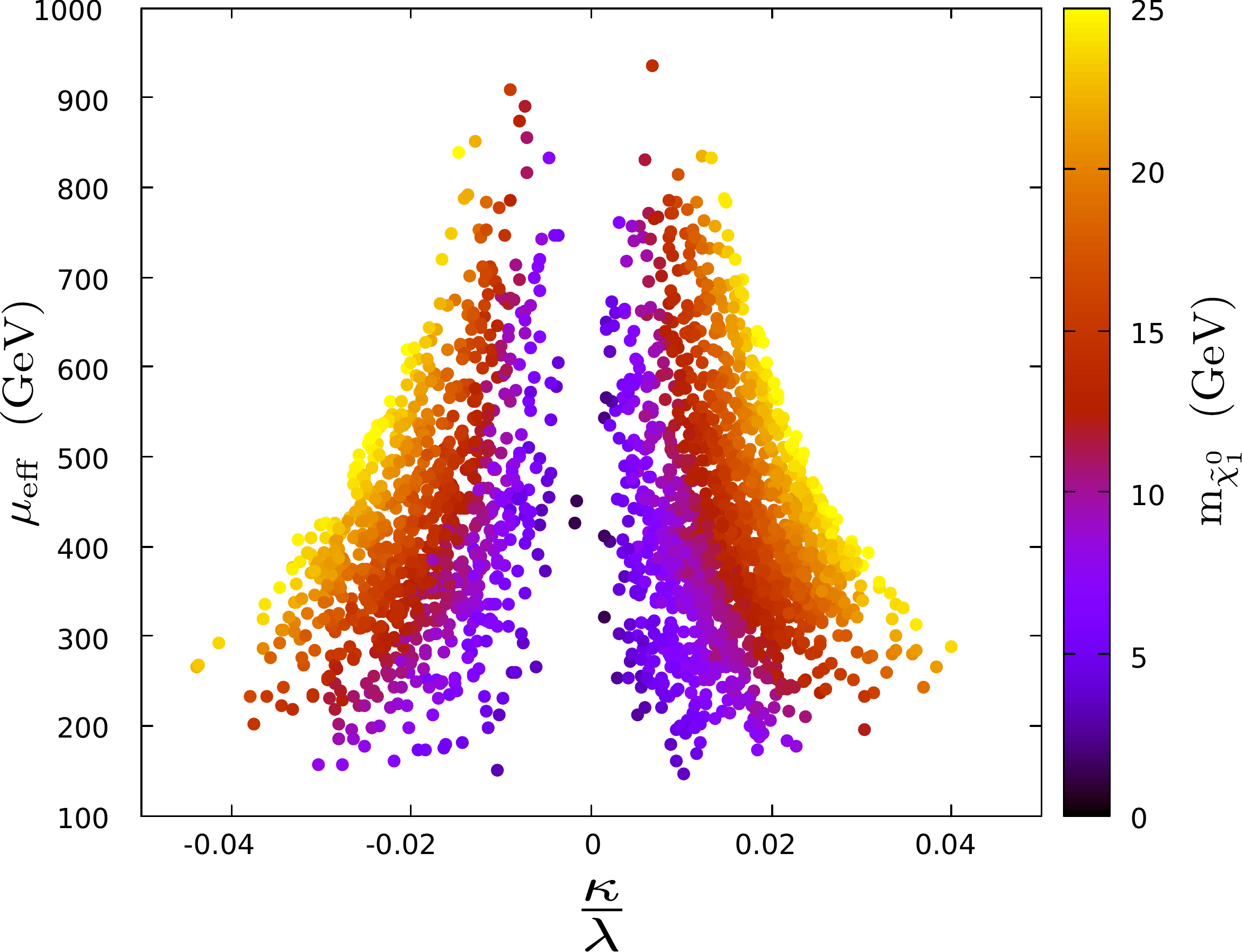}
	\caption{Allowed regions(dotted) in the 
		$\muf$ and $\frac{\kappa}{\lambda}$ plane with the $\rm m_{\N0_1}$.}
	\label{fig:chi1}
\end{figure}

Probable regions of parameters interesting to us are identified performing a naive numerical scan using NMSSMTools\cite{Ellwanger:2004xm,Ellwanger:2005dv}, interfaced with micrOMEGAs\cite{Belanger:2005kh,Belanger:2004yn,Belanger:2006is,Belanger:2013oya} for calculation of DM observables. For the random scan, the numerical ranges of six sensitive parameters (Eq.\ref{eq:param}) are set as:
\br
\rm 0.1 \leq \lambda \leq 0.65,\;  -0.01 \leq \kappa \leq 0.01,\; 1.5 \leq \rm{tan\beta} \leq 20, 100~GeV\leq \mu_{eff}\leq 1000~GeV,\nonumber\\
\rm 500~GeV\leq A_{\lambda} \leq 3500~GeV,\; -100~GeV \leq A_{\kappa} \leq 100~GeV.\; \;\;\;\;\;\;\;\;\;\;
\label{eq:nmssmpara}	
\er
We first performed a scan for a very wide range of these set of parameters, 
and then focus only on the above narrow range which is relevant to the 
signal phenomenology to be studied in this paper. The A-term for third generation($\rm A_t$) plays an important role in predicting the mass of the SM 
like Higgs boson \cite{Hall:2011aa,Ellwanger:2009dp} and is varied for a wider range,
\br
\rm{-8\; TeV < A_{t} < +\, 8~ TeV},
\er
while setting other 3rd generation trilinear parameters as,
\br
\rm{A_{b} = 2\, TeV\; and\; A_{E_{3}} = 1.5\, TeV}.
\er
In order to reduce the number of parameters to vary, all soft masses for left and right handed squarks for the first two generations are assumed as,
\br
\rm{M_{Q_{1,2}}=M_{U_{1,2}}=M_{D_{1,2,3}}=1 \, TeV.}
\er
\br
\rm{1~TeV\leq M_{Q_{3}},\,M_{U_{3}}\leq4\; TeV}
\er
The gaugino masses $\rm{M_{1}}$, $\rm{M_{2}}$ and $\rm{M_{3}}$, which are important for chargino and neutralino sectors are set to be within the range,
\br
\rm{100 \;GeV \leq M_{1} \leq 1 \;TeV,\; 100\; GeV \leq M_{2} \leq 1\; TeV,\; 100\; GeV \leq M_{3} \leq 2\; TeV .}
\er 
Slepton masses of first two generations are fixed to, 
\br
\rm{M_{L_{1,2}}}=300 \;GeV,\; M_{E_{1,2}}=300\; GeV.
\label{eq:MLsoft}
\er
While performing the numerical scan, various constraints, theoretical and as well as experimental, included in NMSSMTools5.5.0\cite{Ellwanger:2004xm,Ellwanger:2005dv} are examined, and accordingly, mass points are rejected or accepted. Precision measurements of the SM-like Higgs boson are used to constrain the model along with the mass requirement of $125 \pm 3$~GeV. In addition, limits on supersymmetric particles obtained at LEP, and Tevatron experiments, and as well as at the LHC are also imposed. Various measurements in flavour physics are also used to check the consistency of mass points. Of course, since lightest neutralino is assumed to be a DM candidate, it is also ensured that the selected mass points are consistent with PLANCK \cite{Aghanim:2018eyx} constraint and Direct searches\cite{Aprile:2018dbl,Akerib:2016vxi,Cui:2017nnn,Agnes:2018ves,Aprile:2019dbj,Amole:2019fdf}. It is to be noted that the numerical (random) scan performed in this study is a representative
one. The main goal of this scan is to identify potential region of parameters interesting to us, and then use few points of the allowed space as benchmark parameters to present the results. Certainly, one needs to perform more rigorous scan in order to find the complete numerical range of parameters through more sophisticated method, e.g those described in Ref.~\cite{Baum:2019uzg,Beskidt:2019mos,Beskidt:2020yvi}.
\begin{figure}[H]
	\begin{center}
		\includegraphics[width=0.5\linewidth]{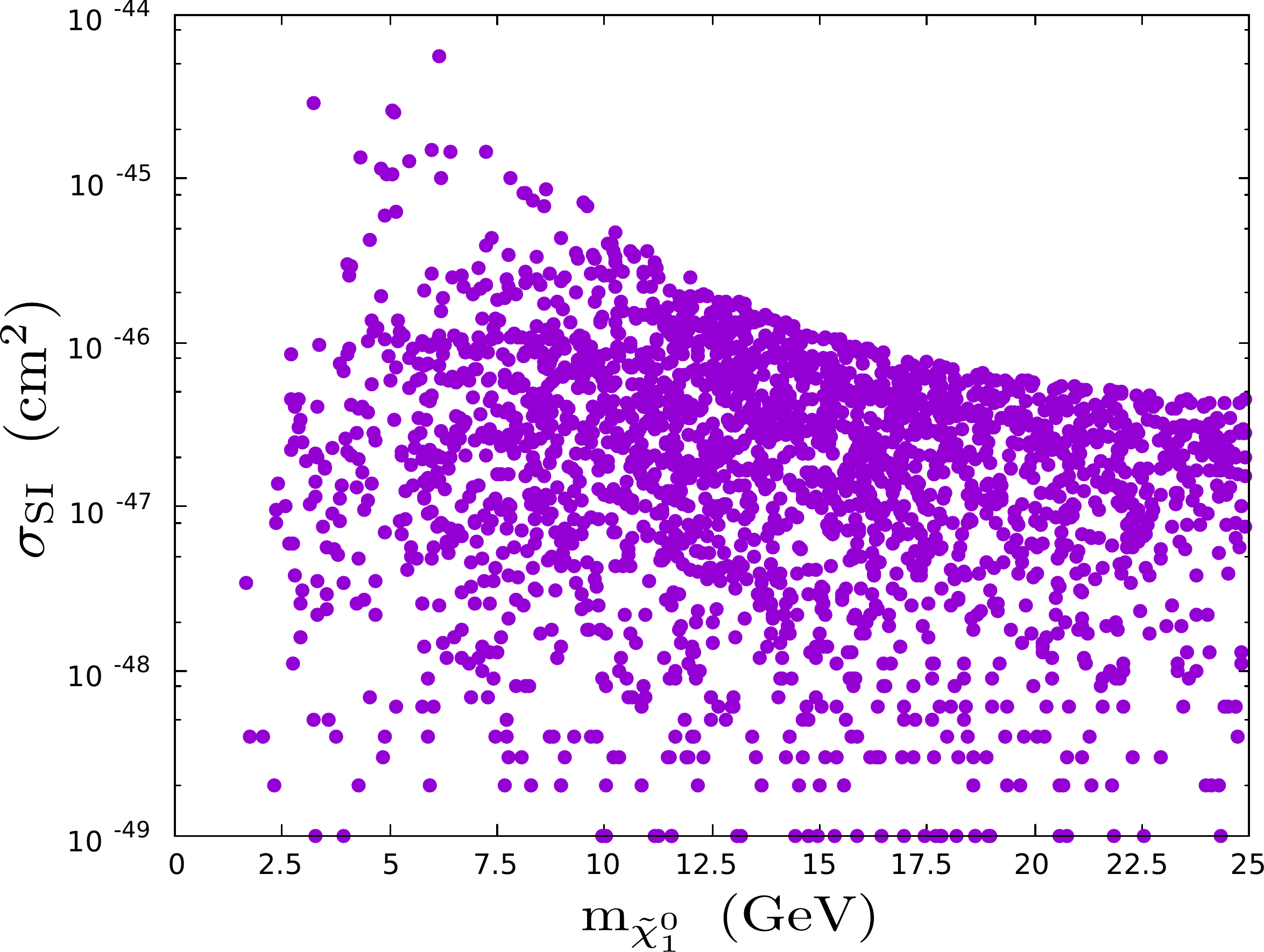}
		\caption{Allowed regions(dotted) of SI DM-nucleon scattering  
			cross-section with the variation of $\N0_1$ masses. 
		}
		\label{fig:dmsdcs}
	\end{center}
\end{figure}
In the following, we present the allowed range of sensitive parameters, 
which are mentioned in the previous section. We focus the region of 
parameters which provide the mass of the lightest singlino-like 
neutralino up to 25 GeV and lightest Higgs bosons almost twice the singlino mass. In Fig.~\ref{fig:dmsdcs}, the spin-independent(SI) DM-nucleon 
cross-sections are presented(dotted) for a range of neutralino masses 
up to 25 GeV and it is also subject to XENON1T and PICO 
constraints~\cite{Amole:2019fdf,Aprile:2018dbl}. It clearly 
demonstrates that the lightest neutralino, even with reasonably low mass, 
can emerge as a viable DM candidate in the NMSSM. In 
Fig.~\ref{fig:chi1}, we show the dependence of lightest neutralino mass corresponding to the interesting range shown in Fig.~\ref{fig:dmsdcs}, on $\frac{\kappa}{\lambda}$ and $\muf$. As anticipated, preferred values are $|\kappa| \sim 10^{-3}$ and $\lambda \sim 10^{-1}$, whereas $\muf \lsim 1~TeV$, not expected to be very large.      

\begin{figure}[H]
	\begin{subfigure}[b]{0.55\textwidth}
		\includegraphics[width=7.7 cm]{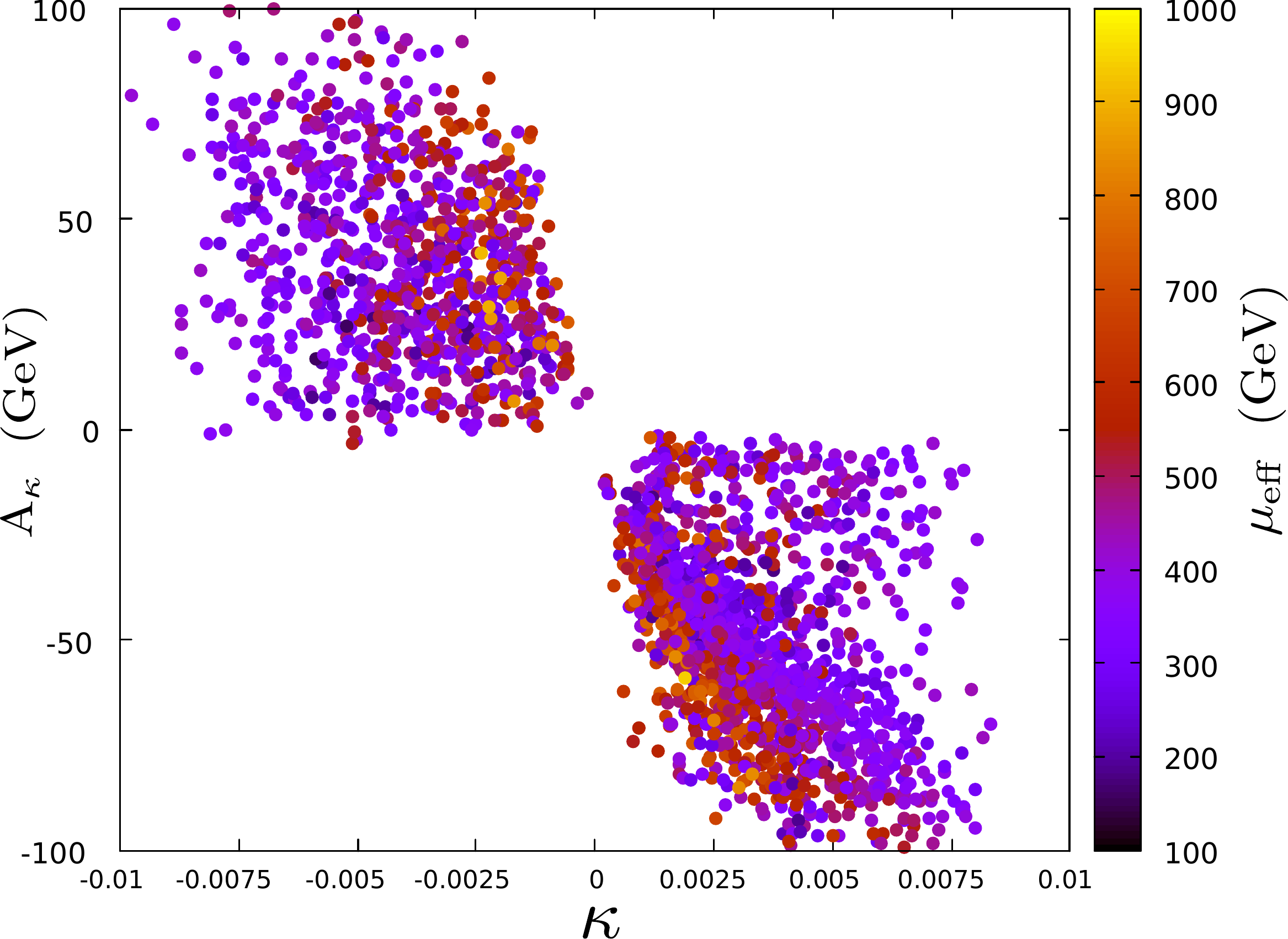}
	\end{subfigure}
	\begin{subfigure}[b]{0.55\textwidth}
		\includegraphics[width=8 cm]{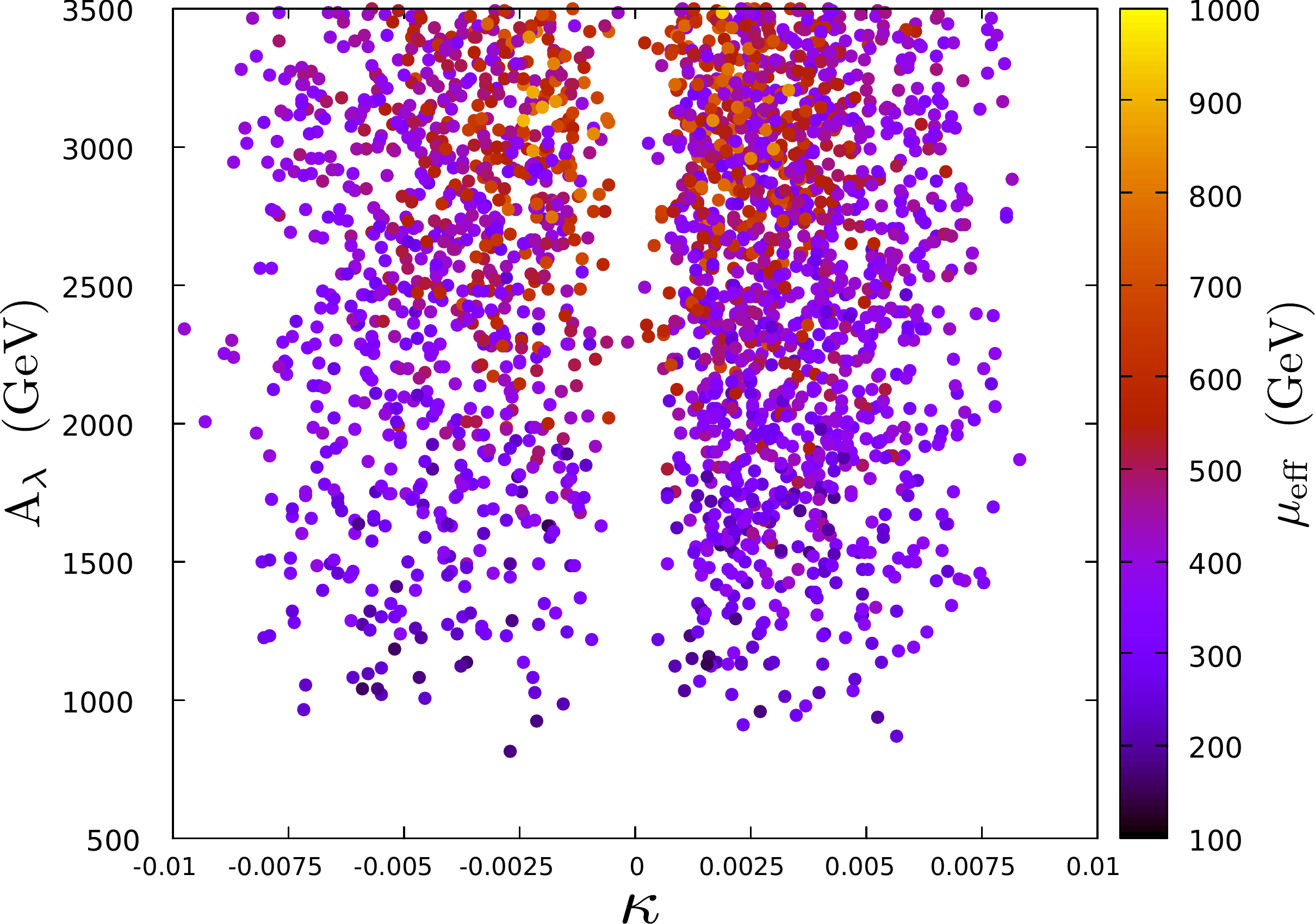}
	\end{subfigure}
	\caption{Allowed regions(dotted) in the $\kappa-A_{\kappa}$ 
		(left) and $\kappa-A_{\lambda}$(right) plane with $\muf$.}
	\label{fig:kAka}
\end{figure}
The tri-linear parameters $\rm A_\kappa$ and $\rm A_\lambda$, play a crucial role along with $\kappa$ and $\lambda$, in determining the masses of Higgs bosons\cite{Miller:2003ay}, in particular, $\rm m_{H_1}$ and $\rm m_{A_1}$. In Fig.~\ref{fig:kAka}, the available region in the $\rm A_{\kappa}-\kappa$ and $\rm A_{\lambda} - \kappa$ plane, relevant to our scenario, are presented along with $\muf$. As argued above, for a very small value of $\kappa$, the large values of $\rm A_{\lambda} \sim {\cal O}(1000)$~GeV and $\rm A_{\kappa}\sim {\cal O}(10)$~GeV are required corresponding to our considered scenario. Value of $|\kappa|\sim 0$ is not permissible and symmetric nature of distribution arises because of the dependence of the value of $\kappa$. We have checked that corresponding to this parameter space (Fig.~\ref{fig:dmsdcs},~\ref{fig:chi1},~\ref{fig:kAka}), the singlet composition in lighter Higgs boson states, and singlino content in lightest neutralino, both are at the level of 95$\%$ or more.

Branching fractions for $\rm H_{SM}\to H_1H_1/A_1A_1$ and 
subsequent decays, $\rm H_1/A_1 \to \N0_1\N0_1 ~ or ~ f\bar f$ 
decide the signal rate. We observe that for a favoured range of 
parameters, such as $\rm \lambda, \kappa, A_{\lambda}$ and 
$\rm A_\kappa$, as discussed above, 
the BR($\rm H_{SM} \to H_1 H_1/A_1 A_1$) $\sim$ 10\% or less, 
which is much below the upper limit 
of $\rm BR(H_{SM}\to BSM)$, constrained by Higgs data, 
and given by~\cite{ATLAS-CONF-2018-031},
\br
\rm BR_{BSM}< 0.26 \;at \;95\% \;C.L.
\label{eq:HBR}
\er       
Branching ratio of light Higgs bosons decay to LSP is also very 
sensitive to $\lambda$ and $\kappa$, as evident 
from Eq.\ref{eq:ghnn1} and \ref{eq:gann1}. A substantial amount of 
singlet composition in light Higgs boson state and singlino content 
in LSP favour this decay channel. However, even a little presence 
of doublet components in light Higgs bosons enhance the 
decay rate in the fermionic channel ($ f \bar f$). Corresponding to our 
interesting region of parameters, the 
BR($\rm H_1/A_1 \to \N0_1\N0_1$) appears to be quite reasonable, and 
sometimes it turns out to be around $\sim$ 70-80\%.

\section{Signal and Background}

In this section, we present the discovery potential of singlino-like 
DM signal at the LHC with the CM energy $\rm \sqrt{s}= 14~TeV$ 
for few luminosity options. We consider the production of light singlet-like Higgs bosons via the non-standard decay channel of the SM Higgs, 
$\rm H_{SM}\to H_1H_1/A_1A_1$, where the mass of $\rm H_1$ or $\rm A_1$ is less than the half of the mass of the SM Higgs boson. Subsequently, 
the lighter Higgs boson states are assumed to decay to lightest  
neutralino pair ($\rm H_1/A_1 \to \N0_1 \N0_1$) with a reasonable 
BR depending on the model parameter space, whereas the other competitive 
decay modes are to heavy fermions, 
like $\rm b\bar{b}$ when kinematically accessible, otherwise $\tau\tau$. 
To ensure harder final state particles, we focus on exclusive 
$\rm H_{SM}+1$ jet process. As we know, the most dominant process of 
Higgs production proceeds via heavy top quark loop leading to, 
$\rm gg \to H_{SM}$~\cite{Heinemeyer:2013tqa,Dittmaier:2011ti}. 
An additional jet originates in next-to-leading order(NLO) perturbative 
QCD with a significant increase of cross-section, either from initial 
gluons or the heavy quarks inside the loop, leading to 
$\rm gg \to H_{SM}+g$. 
Hence, the signal process to our interest appears to be,
\begin{eqnarray}
\rm gg \to \rm H_{SM}+jet \to H_1H_1/A_1A_1 + jet \nonumber\\
\rm \to  b\bar b~{\rm or }~\tau\tau + \N0_1 \N0_1 + jet 
\label{eq:sig} 
\end{eqnarray}
Thus, we focus on signal final state comprising missing energy, which 
is a characteristic of DM signature, along with a reconstructed 
Higgs boson mass accompanied with at least one untagged jet. 

\begin{figure}[H]
	\begin{subfigure}[b]{0.5\textwidth}
		\centering
		\includegraphics[width=7.5cm]{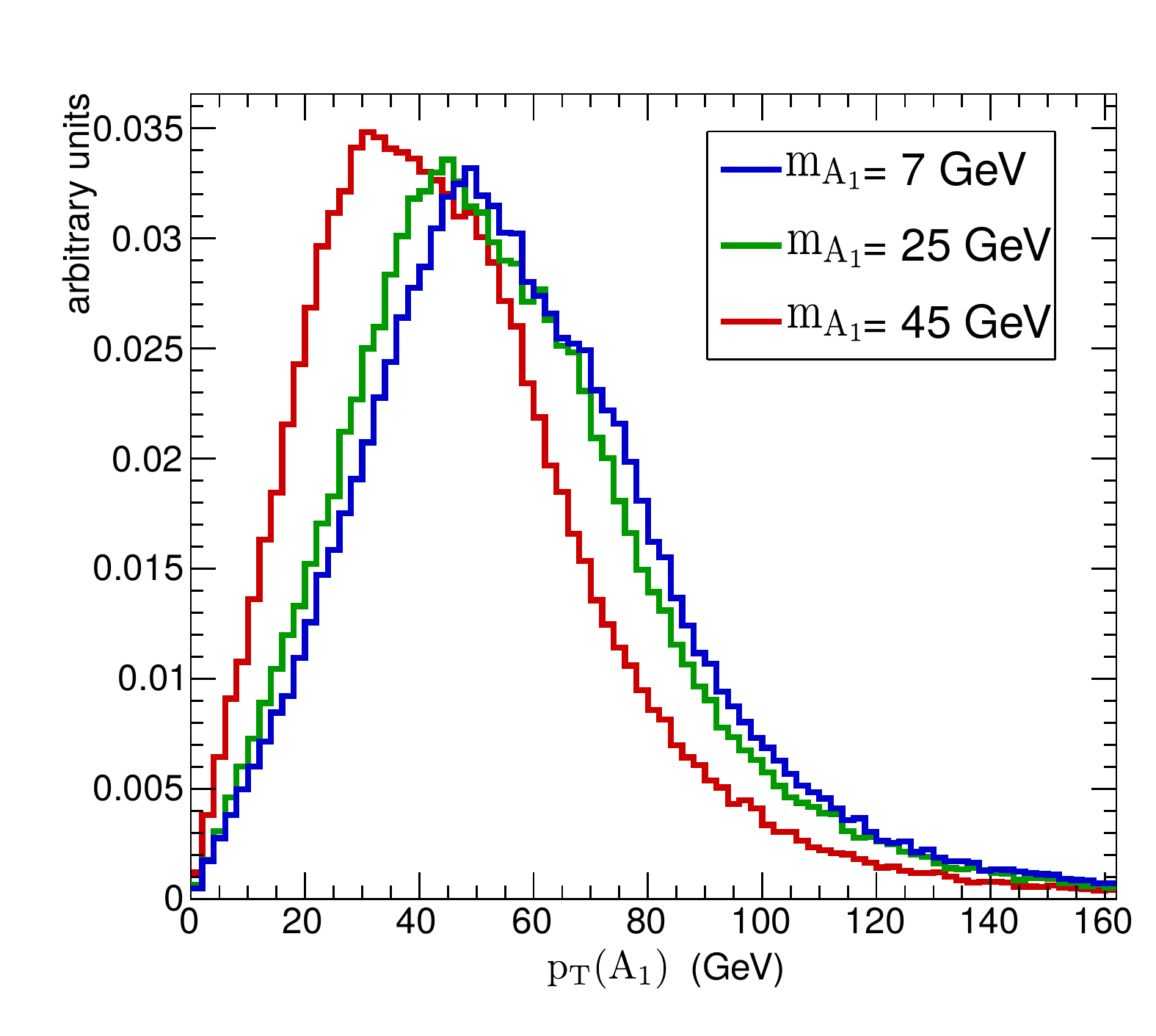}
	\end{subfigure}
	\begin{subfigure}[b]{0.5\textwidth}
		\centering
		\includegraphics[width=7.5cm]{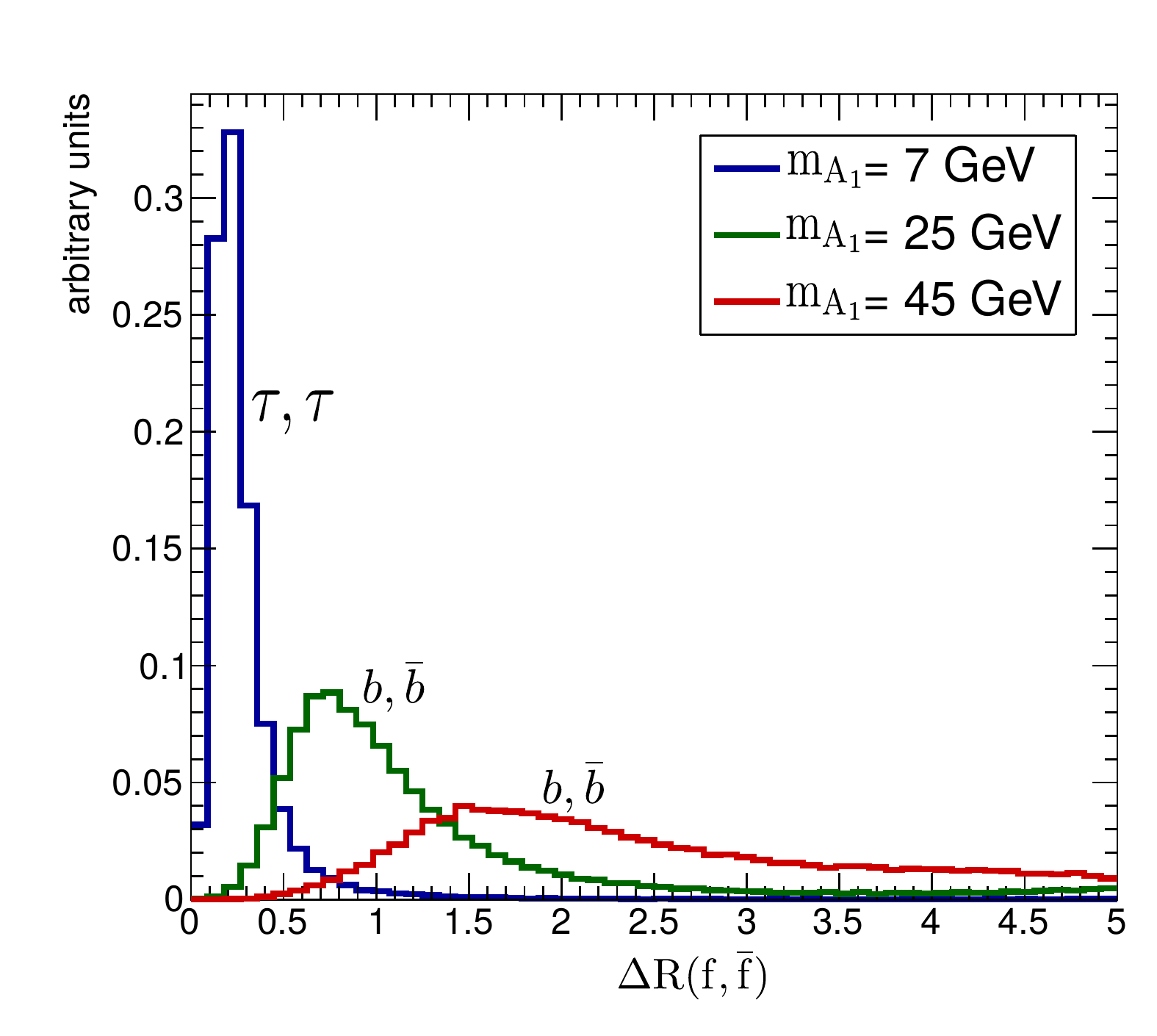}
	\end{subfigure}
	\caption{Transverse momentum of light Higgs boson(left)  
		and $\rm \Delta R$(Eq~\ref{eq:delR})(right) between two fermions 
		originating from the decay of light Higgs bosons.}
	\label{fig:kinem}
\end{figure}

The separation between decay products from lighter Higgs boson is given by \cite{Butterworth:2008iy},
\br
\rm \Delta R(f,\bar f) \simeq \frac{m_{A_{1}/H_1}}{z(z-1)p_T},
\label{eq:delR}
\er                   
implying they are collimated for larger $\rm p_T$ and/or lower mass of parent particle, where 
z is the fraction of momentum of Higgs boson carried by one of the decay 
product. In Fig.~\ref{fig:kinem}, we demonstrate the transverse 
momentum of lighter Higgs boson originating from SM Higgs decay (left) 
and the separation (Eq.~\ref{eq:delR}) between their decay 
products (right) for three sets of Higgs boson 
masses.  

Clearly, the lighter states are more boosted and their decay products are more collimated than those from higher states. These characteristic kinematic features are exploited in 
simulation to isolate signal. Armed with this observation, simulation 
is performed for the signal setting three ranges of the mass 
of $\rm H_1$ or $\rm A_1$, as:
i)  lower mass region: $\rm  m_{H_1/A_1}\leq 10 ~GeV$,
ii) moderate mass region: $\rm 10 ~GeV\leq m_{H_1/A_1}\leq 30 ~GeV$,
and iii) higher mass region: $\rm 30 ~GeV\leq m_{H_1/A_1}\leq 60 ~GeV$.

Notably, as stated above, for `low' and `moderate' mass regions, 
the decay products, either $\rm \tau\tau$ or $\rm b\bar b$ pair 
appears to be very collimated, and emerge as a single `Higgs jet'(HJ) 
with constituents either two  b-like($\rm J_{b\bar{b}}$)  or $\tau$ ($\rm J_{\tau{\tau}}$) -like subjets 
depending on the decay modes. 
Hence, instead of tagging individual $\tau$-jet or b-jet, 
which is challenging in this present scenario, 
`Higgs jet' is tagged to classify signal from the background.  
On the contrary, tagging HJ is not very effective for ``high mass region'',
since decay products emerge with a wider separation. In this case, 
we observed that even losing signal events due to tagging of HJ, 
still it is very useful to reduce the SM backgrounds 
substantially. Hence, in summary, simulation is performed for 
three categories:
\br
\rm{J_{\tau\tau}+\MET+\geq 1 \,j\;\;\; for\;m_{H_1/A_1}\leq 10 ~GeV \;},
\label{eq:signal0}
\er
\br
\rm{J_{b\bar{b}}+\MET+\geq 1 \,j \;\;\; for\; 10 ~GeV\leq m_{H_1/A_1}\leq 
	30 ~GeV\;},
\label{eq:signal1}
\er
\br
\rm{J_{b\bar{b}}+\MET+\geq 1 \,j \;\;\; for\; 30 ~GeV\leq m_{H_1/A_1}\leq 
	60 ~GeV\;}.      
\label{eq:signal2}       
\er
We discuss signal selection strategy for the case of lower mass range, 
Eq. ~\ref{eq:signal0}, in a later subsection separately.

The dominant sources of SM backgrounds corresponding to the signal 
processes(Eq.~\ref{eq:signal1} and \ref{eq:signal2}) are due to 
the processes: 
\br
\rm p p \rightarrow \rm{ t\bar{t},\; Wb\bar{b}+jets},\; 
Zb\bar{b}+jets,
\label{eq:Bg}
\er 
Neutrinos originating from W or Z decay contribute to missing transverse 
energy ($\rm \MET$). We also checked the level of background contribution from 
WZj, ZZj, $\rm H_{SM}Wj$ and $\rm H_{SM}Zj\; (H_{SM}\sim H_2)$ and found to be very small due to comparatively very low cross-sections and respective branching ratios.

For the sake of illustration, six benchmark points (BP), as shown 
in Table~\ref{BPtable}, compatible with various experimental data, 
are chosen to simulate the signal process. These BPs are selected such 
that $\rm 2 m_{\N0_1} \sim m_{H_1/A_1}$ and covering mass ranges as 
required in Eq.~\ref{eq:signal0} -- \ref{eq:signal2}. Notice that for 
all such cases, $\rm H_2$ turns out to be the SM-like Higgs boson and 
decays to a pair of non-SM-like Higgs bosons states 
$\rm H_2 \to H_1H_1 / A_1A_1$, with a BR ranging from $\sim$ 
0.01\% to 10\%, which is within the constraint 
given by (Eq.~\ref{eq:HBR}). 
As mentioned before, light Higgs bosons, mainly decay to 
either in $\rm b\bar b$ or $\N0_1\N0_1$ channel, which we require 
for our signal process.     

\begin{table}[H]
	\caption{Parameters, BRs, Higgsino components($\rm N_{13}^2, ~N_{14}^2$) of the singlino and fraction of annihilation channels 
		contributing to relic density 
		corresponding to few benchmark points(BP). Energy units are in GeV.}
	\resizebox{\textwidth}{!}{
\begin{tabular}{c c c c c c c}
	\hline
	 &BP1 & BP2 & BP3 & BP4 & BP5 & BP6\\ 
	\hline
	$\lambda$ & 0.34195 & 0.17783 & 0.22140 & 0.24670 & 0.24980 & 0.29853\\
	  
	$\kappa$ & 0.00080 & 0.00241 & -0.00564 & 0.00520 & -0.00690  & 0.00438\\
	
	$\rm tan\beta$ & 8.46 & 5.99 & 4.79 & 5.85 & 4.96 &  4.63\\ 
	
	$\rm A_{\lambda}$ & 3114.53 & 793.52 & 1201.50 & 1654.39 & 1968.95 & 1528.60\\
	
	$\rm A_{\kappa}$ &  -46.48 & -29.91 & 36.66 & -57.21 & 69.65 & -60.15\\
	
	$\muf$  & 340.39 & 150.68  & 232.94 & 290.40  & 378.55 & 364.86\\
	\hline
	$\rm m_{H_{2}}$ & 123 & 126  &  126 & 126  &  123 & 127\\
	
	$\rm m_{H_{1}}$ & 43 & 14  &  28 & 36  &  44 & 56\\
	
	$\rm m_{A_{1}}$ & 8 & 12  &  24 & 31  &  47 & 30\\
	
	$\rm m_{\tilde{\chi}_{1}^{0}}$ & 3 & 5  &  10 & 14 &  20 & 13\\
	\hline
	$\rm N_{13}^2$  & $10^{-4}$  & $4\times10^{-7}$  & $3\times10^{-4}$  & $10^{-6}$  & $3\times10^{-4}$ & $10^{-4}$\\
	$\rm N_{14}^2$  & 0.03  & 0.04  & 0.02  & 0.02  & 0.01 & 0.02\\
	\hline  
	$\rm \Omega h^{2}$ & 0.1115 & 0.1188  &  0.1188 & 0.1255  &  0.1180 & 0.1098\\
	\hline
	$\rm BR(H_{2}\rightarrow H_{1} H_{1})$ &  0.0001 & 0.06   & 0.01 & 0.11   & 0.08  & 0.07\\
	
	$\rm BR(H_{2}\rightarrow A_{1} A_{1})$ & 0.10 &  0.004  & 0.06 &  0.001  &  0.02 & 0.01 \\
	
	$\rm BR(H_{1}\rightarrow b \bar{b})$ & 0.81  & 0.57   & 0.75 & 0.22   & 0.50 & 0.50\\
	$\rm BR(H_{1}\rightarrow \tilde{\chi}_{1}^{0} \tilde{\chi}_{1}^{0})$ & 0.07  & 0.31 & 0.18 & 0.75  & 0.45 & 0.44\\
	$\rm BR(H_{1}\rightarrow \tau \tau)$   & 0.07 &  0.08 &  0.06 &  0.02  &  0.04 & 0.05\\
	
	$\rm BR(A_{1}\rightarrow b\bar{b})$ & --  &  0.35 &  0.32 &  0.55   & 0.18 & 0.73\\ 
	$\rm BR(A_{1}\rightarrow \tilde{\chi}_{1}^{0} \tilde{\chi}_{1}^{0})$ & 0.22 & 0.13  & 0.64 & 0.40  & 0.80 & 0.19 \\
	$\rm BR(A_{1}\rightarrow \tau \tau)$  & 0.69 &  0.42 &  0.03  & 0.05  & 0.01   & 0.06 \\
	
	\hline
	Annihilation  & 0.90 ($\rm \tau$$\tau$) & 0.90 ($\rm b\bar{b}$)  & 0.92 ($\rm b\bar{b}$) & 0.92 ($\rm b\bar{b}$)  & 0.91 ($\rm b\bar{b}$) & 0.92($\rm b\bar{b}$)\\
	channels  & 0.09 (gg) & 0.07 ($\rm \tau$$\tau$)  &  0.07 ($\rm \tau$$\tau$) & 0.07 ($\rm \tau$$\tau$)  &  0.08 ($\rm \tau$$\tau$) & 0.07 ($\rm \tau$$\tau$)\\
	& & 0.02 (gg) & 0.01 ($\rm c \bar{c}$) & & & \\
	\hline
\end{tabular}
}
	\label{BPtable}
\end{table}
PYTHIA8 is used to simulate $\rm t \bar t$ events, while other processes are generated using 
Madgraph5-aMC@NLO-2.6.4\cite{Alwall:2014hca} and 
PYTHIA8 \cite{Sjostrand:2006za,Sjostrand:2007gs} for subsequent 
showering and hadronization. The signal events are generated using 
PYTHIA8 inputting masses and branching ratios of SUSY particles 
and Higgs bosons through SLHA file \cite{Skands:2003cj} which is 
generated using NMSSMTools. To take into account detector effects, generated events for both signal and backgrounds are passed through Delphes-3.4.2\cite{deFavereau:2013fsa} using the CMS detector card. The Delphes objects, namely, eflows are used for analysis.

In the simulation, events are selected adopting the following strategy.
\begin{itemize}
	\item
	Lepton veto: Events consisting leptons are vetoed out. Leptons are selected with $\rm p_T^\ell>10$~GeV and $\rm |\eta|<$2.5. It reduces the background events significantly without losing any signal.
	\item
	HJ selection: The e-flow objects (e-flow tracks, e-flow photons and e-flow 
	neutral hadrons) of Delphes are given as input to 
	Fastjet3.3.2 \cite{Cacciari:2011ma} to construct fat jets. The 
	Cambridge-Aachen \cite{Dokshitzer:1997in} algorithm is used setting 
	the jet size parameter R=1 and 1.6 for moderate and high mass 
	regions (Eq.\ref{eq:signal1} and \ref{eq:signal2}) of lighter Higgs bosons respectively. 
	The Fatjets are selected with $\rm p_T^J>$40~GeV and $|\eta|<$4.0. Mass-drop Tagger (MDT)\cite{Butterworth:2008iy,Dasgupta:2013ihk} with $\rm \mu=$0.667 and $\rm y_{cut}>$0.01 is used to tag Fat-jets with two subjets.  
	The subjets of `tagged fat jet' are further matched with the b-quarks of the events which are selected with a minimum $\rm p_T$ cut of 0.5 GeV and $|\eta|<2.5$ with a matching cone $\rm \Delta R <$0.3, where $\rm \Delta R = \sqrt{(\eta_{q}-\eta_{j})^{2}+ (\phi_{q}-\phi_{j})^{2}}$; $\rm \eta_{q}$, $\rm \eta_{j}$ are pseudorapidities and $\rm \phi_{q}$, $\rm \phi_{j}$ are azimuthal angles of b-quark and jet respectively. If both of the sub-jets are found to be b-like satisfying matching criteria, then it is claimed to be tagged as the HJ ($\rm J_{b\bar b}$). We found that the tagging efficiency of $\rm J_{b\bar b}$ is around 30\% for the lower range of light Higgs boson mass and goes down to around 15\% for higher mass range. The mass of $\rm J_{b\bar b}$ is depicted in Fig.~\ref{fig:mbb} for three 
	samples of Higgs bosons masses. Clearly, the mass peaks are observed at 
	the given input masses. However, peaks are observed to be broader 
	for higher Higgs boson masses. In the same figure, the corresponding 
	distributions from backgrounds are also shown, which are not showing 
	clearly any peaks, as expected. Notably, the presence 
	of $\rm J_{b\bar b}$ with a peak in its mass
	distribution is the characteristics of our signal events. 
	\item
	Non-tagged jets: After tagging $\rm J_{b\bar b}$, non-tagged QCD jets are constructed out of remaining hadrons in the events using 
	Anti-$\rm k_T$\cite{Cacciari:2008gp} algorithm with a jet size parameter 
	R=0.5. The reconstructed jets are selected with $\rm p_{T}^{j}>$20 GeV and
	$|\eta|<$4.0.
	\item
	Missing transverse momentum($\rm \MET$): The missing transverse momentum is 
	constructed  by vector addition of momenta of all visible particles,
	i.e. ${\vec{\rm p_{T}}}= - \sum \vec{\rm p_{T}^{i}}$, where i runs over all 
	constructed collection from the Detector. Delphes stores $\rm \MET$ of each 
	events taking into account detector effects.
\end{itemize} 

\subsection{Signal for low mass of $\rm H_1/A_1$}
In this sub-section, we discuss the search strategy of the signal process, 
Eq.~\ref{eq:signal0}, which is very challenging since the masses of intermediate Higgs bosons are too low to have energetic decay products. The decay mode of Higgs bosons to a pair of $\tau$ leptons is preferred over the $\rm b \bar b$ in order to avoid huge QCD background, decay channel and for the same reason, the hadronic mode of tau 
leptons leading to $\tau$-jets are not considered. 
Hence, in this scenario, we focus on the final 
state following Eq.~\ref{eq:signal0} as,
\br
\ell^+ \ell^- +\rm \MET+\geq 1~jet .
\label{eq:lowmass}
\er   
Note that the combined BR for both the $\tau$ leptons 
decaying leptonically is very 
small ($\sim$ 12\%). Moreover, leptons are too soft with a very low  
$\rm p_T \sim \frac{m_{H_1/A_1}}{2}$. In this case, dominating sources of SM 
backgrounds are due to the inclusive Drell-Yan, ${\rm t \bar t}$, and electroweak processes
W+jets, WW+jets, WZ+jets. 
Performing a naive simulation for both signal and background, 
we try to find the signal sensitivity.  
For all background processes except $\rm t\bar t$, matrix elements are generated in MadGraph5aMC@NLO-2.6.4(MG5NLO), then showering and hadronization are performed 
using PYTHIA8 as before. The $\rm t \bar t$ events are fully generated 
using PYTHIA8. In the simulation, 
leptons(both $e$ and $\mu$) are selected  with $\rm p_T^\ell \ge$~10~GeV 
and $|\eta|<$2.5. {\footnote{Experimentally lepton trigger of
		low $\rm p_T$ are to be used}}
Requirement of isolated leptons reduces the signal event significantly.  
The two leptons originating from $\tau$ pairs are not 
expected to be widely separated. In our simulation, we ensure 
isolated leptons by checking e-flow objects of Delphes 
using following criteria as,
\br
\rm \frac{\sum p_{T}^{R<0.2}}{p_{T,\ell}}<0.1 , \;\; \ell=e,\mu
\er     
where $\rm p_T^{R<0.2}$ is the sum of the transverse momentum of all particles 
which are within $\rm \Delta R<0.2$ with respect to lepton momentum 
direction. It also ensures that both the signal leptons are separated
by $\rm \Delta R>$0.2.  
Construction of $\rm \MET$ and jets (including b-jets) are the same as 
before and performed by Delphes.

\section{Results and Discussion} 
\begin{figure}[H]
	\centering
	\includegraphics[width=0.5\linewidth]{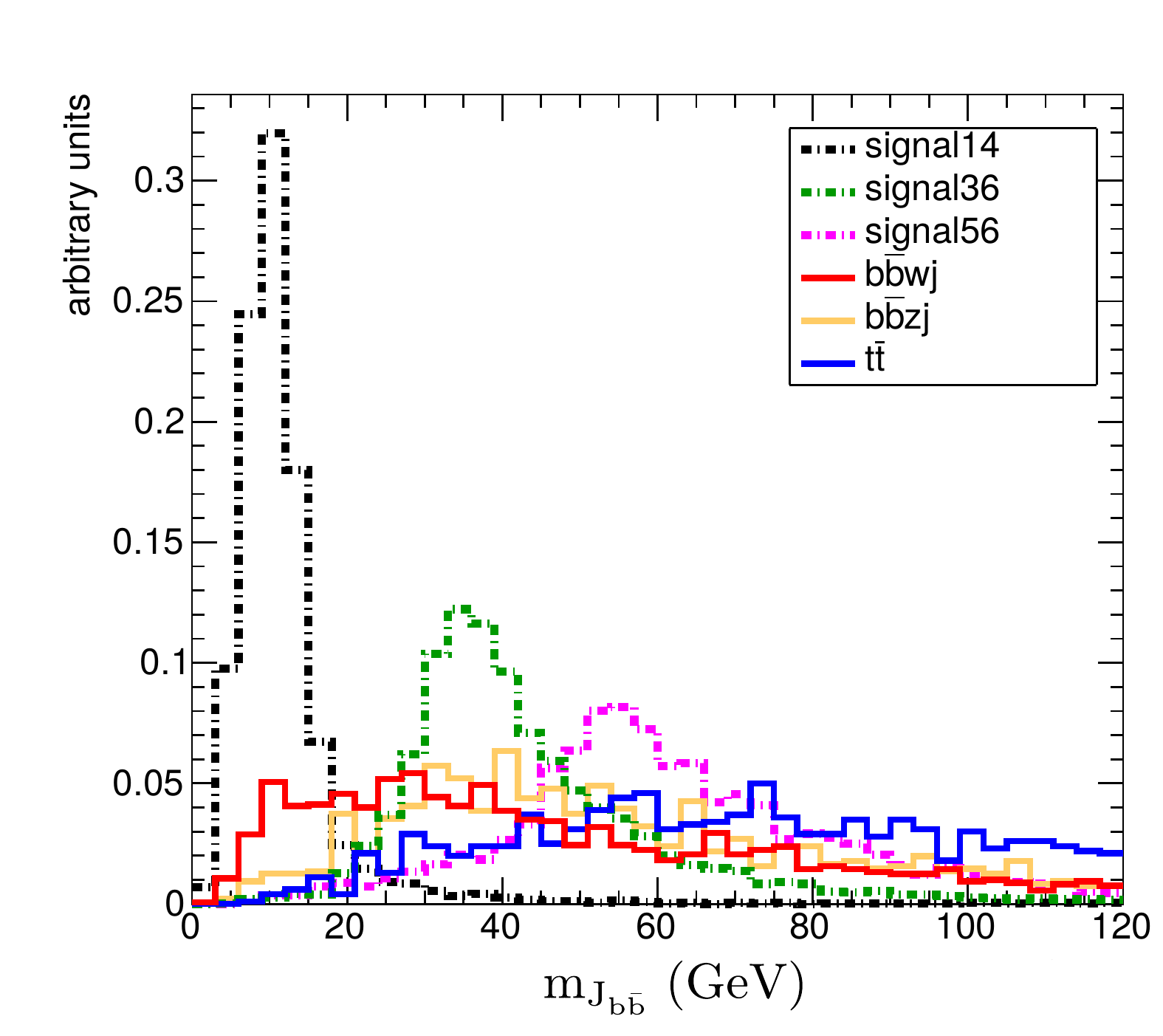}
	\caption{Mass of $\rm J_{b\bar b}$ for three signal points ($\rm m_{H_1}=$14, 36 and 56 GeV), and dominant sources of 
		backgrounds.}
	\label{fig:mbb}
\end{figure}

Identifying various distinguishing features of the signal process, we impose 
a few event selection cuts to eliminate backgrounds. For example, the characteristics of $\rm J_{b\bar b}$ mass ($\rm m_{J_{b\bar{b}}}$) distribution, as shown in Fig.~\ref{fig:mbb}, are very different for backgrounds and signal events. 
Therefore, a background rejection cut setting as,
\br
\rm m_{J_{b\bar{b}}} &<& 30~ \rm {GeV~for~lower~mass~range},\nonumber \\
\rm 30 &<& \rm  m_{J_{b\bar{b}}} < 60~ \rm {GeV~for~higher~mass~range},
\label{eq:mjbb}
\er
is highly effective, in particular, for eliminating $\rm t\bar t$ background 
by 70-80\%.

\begin{figure}[H]
	\begin{subfigure}[b]{0.51\textwidth}
		\centering
		\includegraphics[width=8cm]{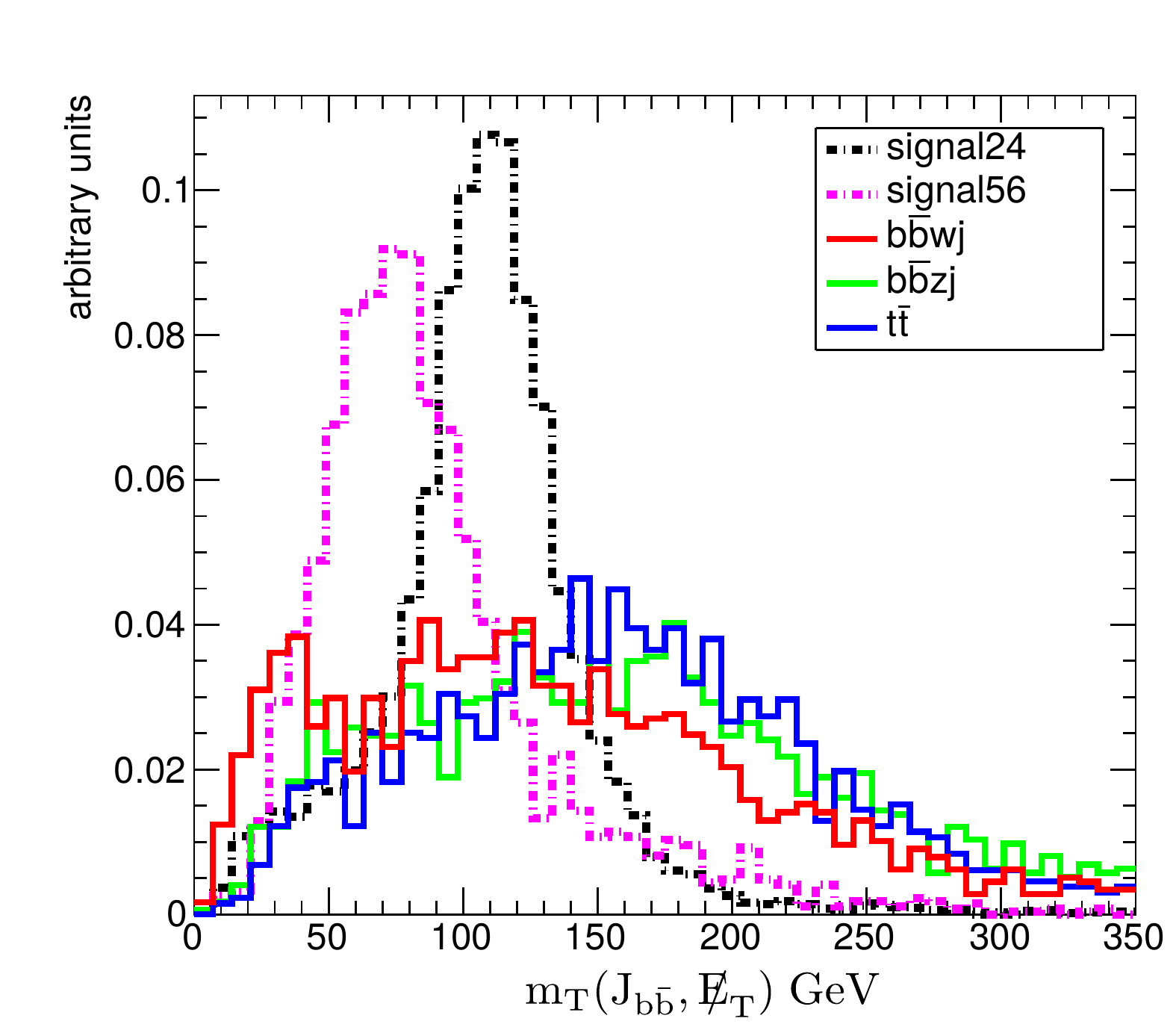}
	\end{subfigure}
	\begin{subfigure}[b]{0.5\textwidth}
		\centering
		\includegraphics[width=8 cm]{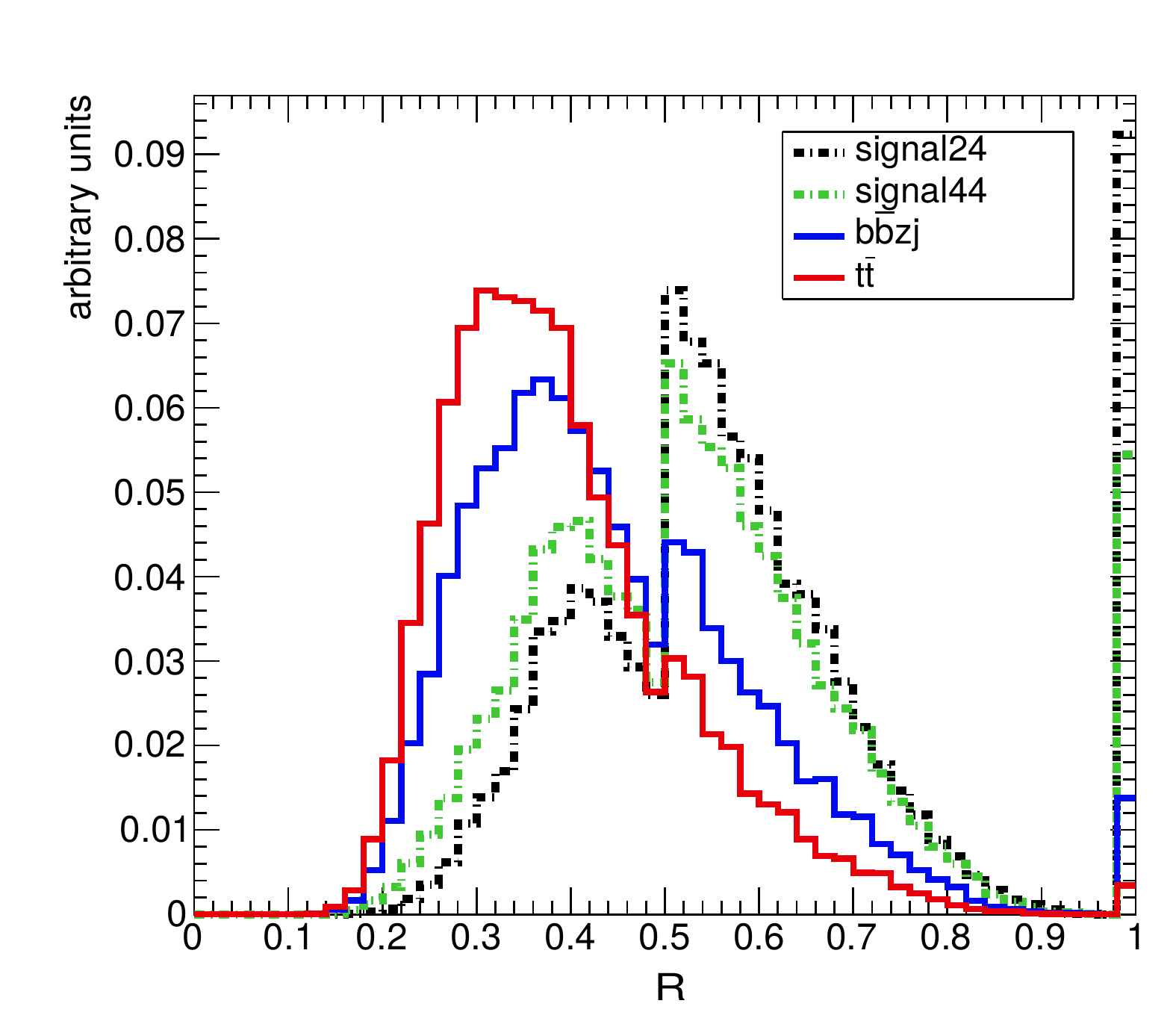}
	\end{subfigure}
	\caption{Transverse mass between $\rm J_{b\bar{b}}$ and $\rm \MET$(Eq.~\ref{eq:mtbb}) (left) 
		and R(Eq.\ref{eq:rt})(right) for signal (with $\rm m_{A_1}=$24 GeV and $\rm m_{H_1}=$56 (left) or 44 (right) GeV), 
		$\rm b\bar{b}Z+jets$ and $\rm t\bar{t}$.}
	\label{fig:MTRT}
\end{figure}
Evidently, the transverse mass between $\rm J_{b\bar b}$ and $\rm \MET$ is restricted by the SM Higgs boson mass in signal, as shown in Fig.~\ref{fig:MTRT} (left), which is not the case for backgrounds. Hence an upper cut on it as,
\br
\rm m_T(J_{b\bar{b}}, \MET) = \sqrt{2\times p_T^{J_{b\bar{b}}}\times \MET\times (1-\cos\phi(J_{b\bar{b}}, \MET))} < 140 ~GeV, 
\label{eq:mtbb}
\er 
is found to be helpful in suppressing background. Another interesting observable is useful in reducing the top background, which is defined as\cite{Guchait:2011fb},
\br
\textrm R(n_{j}^{min})=\frac{\sum_{i=1}^{n_{j}^{min}} 
	|\vec{p_{T}}^{j_{i}}|}{\rm H_{T}},
\label{eq:rt}
\er 
where $\rm n_j^{min}$ is the minimum number of jets required in event 
selection and $\rm H_{T}$ $=\sum_{i=1}^{n_{j}}|\vec{p_{T}}^{j_{i}}|$. 
Obviously, by construction $\rm 0<R\leq 1$, 
where $\rm n_j^{min}$ is set equal to 1 for signal event selection. 

Distribution of R is expected to be on higher side ($\rm R\sim 1$) for signal,
since it is not very jetty, whereas for $\rm t\bar t$ it is expected to be on lower side, as shown in Fig.~\ref{fig:MTRT}(right). Therefore, a selection on $\rm R>$0.5 suppresses a good fraction of top events and to some extent
$\rm Zb\bar{b}+jets$ events for moderate mass region. 

Cross-section yields for the signal which are subject 
to two different sets of cuts(Eq.~\ref{eq:mjbb}) on $\rm m_{J_{b\bar{b}}}$ corresponding to benchmark 
points, and background processes after each set of cuts are 
presented in Tables~\ref{tab:lmevt} and \ref{tab:hmevt}.

The first row presents the leading order(LO) cross-sections with the center 
of mass energy $\sqrt{s}=14$~TeV, setting  NNPDF23LO \cite{Ball:2010de} 
for parton distribution and choosing the dynamic scale ($\rm \sqrt{m^2+p_T^2}$) 
computed by Madgraph5-aMC@NLO-2.6.4\cite{Alwall:2014hca}. 
Cross-sections for the background processes($\rm Zb\bar{b}+jets$, 
$\rm Wb\bar{b}+jets$) are computed in Madgraph5-aMC@NLO-2.6.4 
in five flavour scheme and subject to cuts, 
$\rm p_T^b >$20~GeV, $\rm p_T^j>$20~GeV, 
$\rm \Delta R(b,b)>$0.1 and $\rm \Delta R(j,j)>$0.4. Higher order 
effects to all these cross-sections are taken into account 
through K-factors, as defined, $\rm K=\frac{\sigma_{NLO}}{\sigma_{LO}}$. 
These K-factors are obtained by computing respective cross-sections 
using MCFM\cite{Campbell:1999ah,Campbell:2011bn,Campbell:2015qma,Boughezal:2016wmq}. 
The K-factors of the processes, $\rm Zb\bar{b}+jets$ and $\rm Wb\bar{b}+jets$, 
are considered to be the same as for the 
processes $\rm Zb\bar{b}$ and $\rm Wb\bar{b}$, which are computed by MCFM and found to be  $\sim$1.7 and $\sim$2.6 respectively, and in close 
agreement with Ref\cite{Cordero:2009kv}. For $\rm t\bar t$, K-factor=1.4 
is used~\cite{Melnikov:2009dn,Kidonakis:2008mu}. For signal process, K-factor is estimated to be $\sim$ 
1.8 using MCFM, close to quoted values in 
Ref.~\cite{deFlorian:2016spz}. 
All these K-factors are taken into account in 
Tables~\ref{tab:lmevt} and \ref{tab:hmevt} 
while presenting final yields at the end. In these tables, $\rm \epsilon_{BR}$ is the sum of the branching ratios $\rm BR(H_{2}\rightarrow H_{1} H_{1})$ and 	$\rm BR(H_{2}\rightarrow A_{1} A_{1})$.

Events are required to contain at least one jet with cuts $\rm p_T^j>$20~GeV 
and $|\eta|<3$ and vetoed out if there be any lepton. The $\rm \MET$ cut 
is useful in reducing the backgrounds, in particular due to the process 
with a Z and W boson in the final state, however, it costs 
signal also by almost a factor of 2, even it is more severe 
for signal corresponding to lower mass $\sim$15 GeV.
Notice that the selection of $\rm J_{b\bar b}$, and the respective mass 
window (Eq.~\ref{eq:mjbb}) suppress backgrounds substantially, by almost 
two orders of magnitude, while signal remains less affected.
A cut on the transverse mass, Eq.~\ref{eq:mtbb}, is very effective in 
isolating the backgrounds without costing signal events 
too much, as seen in both the tables. 
Eventually, as expected, the cut on R
suppresses the top background further by about $\sim$ 50\%. 

\begin{table}[H]
	\caption{Cross-section yields after each set of cuts 
		for two low mass signal points BP2 and BP3(Table~\ref{BPtable}) 
		and background processes. Last row presents the
		final cross-sections after including K-factor and b tagging efficiency.}
	\centering
	\begin{tabular}{c c c c c c }
 \hline   & BP2  & BP3  & $\rm b\bar{b}Z+jets$  & $\rm b\bar{b}W+jets$  & $\rm t\bar{t}$  \\ 
 \hline $\sigma$(pb)  & 12.4  & 12.4  & 152.8  & 139.8  & 597.9  \\ 
 $\rm \sigma\times \epsilon_{BR}$ & 0.7 & 0.9 & 152.8  & 139.8 &  597.9\\
 lepton veto  & {$0.6$}  & {$0.8$}  & {$108.5$}  & {$97.6$} & {$298.2$}  \\ 
 $\rm n_{j}\geq1$  & {$0.5$}  & {$0.7$}  & {$107.4$}  & {$96.3$}  &  {$297.7$}  \\ 
 $\rm \MET>40.0 ~GeV$  & {$0.3$}  & {$0.4$}  & {$32.8$}  & {$24.4$}  &  {$109.4$}  \\ 
 $\rm No.\; of \;J_{b\bar{b}}$ =1  & {$0.05$}  & {$0.06$}  & {$1.8$}  & {$3.0$}  & {$4.9$}  \\ 
 $\rm  m_{J_{b \bar{b}}}<30.0$ GeV  & {$0.05$}  & {$0.05$}  & {$0.3$}  & {$1.0$}  & {$1.3$}  \\ 
 $\rm  m_{T}(J_{b\bar{b}},\MET) \leq $140 ~GeV  & {$0.04$}  & {$0.04$}  & {$0.2$}  & {$0.8$}  & {$0.9$}  \\ 
 R$>$0.5  & {$0.034$}  & {$0.04$}  & {$0.08$}  & {$0.6$}  & {$0.4$}  \\ 
 $\rm \sigma$$\times$ K-factor$\rm \times \epsilon_{b}^2$ & 0.018 & 0.022 & 0.04 & 0.47 & 0.24\\
\hline \end{tabular}

	\label{tab:lmevt}
\end{table}

\begin{table}[H]
	\caption{Same as Table~\ref{tab:lmevt}, 
		but for three `high mass' points, BP4, BP5 and BP6) in Table~\ref{BPtable}.}
	\centering
	\resizebox{\textwidth}{!}{
\begin{tabular}{c c c c c c c }
 \hline   & BP4  & BP5  & BP6  & $\rm b\bar{b}Z+jets$  & $\rm b\bar{b}W+jets$   & $\rm t\bar{t}$  \\ 
 \hline $\sigma$(pb)  & 12.4  & 12.4  & 12.4  & 152.8  & 139.8  & 597.9  \\ 
 $\rm \sigma\times \epsilon_{BR}$ & 1.3 & 1.2 & 1.0 & 152.4  & 139.8 & 597.9\\
 lepton veto  & {$1.3$}  & {$1.1$}  & {$0.9$}  & {$108.6$}  & {$97.6$}  & {$298.2$}  \\ 
 $\rm n_{j}\geq1$  & {$1.2$}  & {$1.0$}  & {$0.9$}  & {$108.0$}  & {$97.3$}  & {$297.8$}  \\ 
 $\rm \MET>35.0 ~GeV$  & {$0.9$}  & {$0.6$}  & {$0.4$}  & {$39.4$}  & {$30.4$}  & {$127.9$}  \\ 
 $\rm No.\; of \;J_{b\bar{b}} =1$  & {$0.05$}  & {$0.04$}  & {$0.03$}  & {$3.0$}  & {$2.9$} & {$7.8$}  \\ 
 $\rm 30.0 <m_{J_{b \bar{b}}}<60.0~ GeV$  & {$0.03$}  & {$0.03$}  & {$0.01$}  & {$0.6$}  & {$0.8$}  & {$1.8$}  \\ 
 $\rm  m_{T}(J_{b\bar{b}},\MET) \leq $140 GeV  & {$0.03$}  & {$0.03$}  & {$0.01$}  & {$0.5$}  & {$0.5$}  & {$1.2$}  \\ 
 R$>$0.5  & {$0.024$}  & {$0.02$}  & {$0.01$}  & {$0.26$}  & {$0.4$}  & {$0.5$}  \\ 
 $\rm \sigma$$\times$ K-factor$\rm \times \epsilon_{b}^2$ & 0.013 & 0.011 & 0.0055 & {$0.13$} & {$0.3$} & 0.3\\
\hline \end{tabular}
}

	\label{tab:hmevt}
\end{table}
Finally, in order to obtain final cross-section yields, we take into account 
$\rm p_T$-dependent b-tagging efficiency ($\rm \epsilon_b$)~\cite{Sirunyan:2017ezt}. For $\rm t\bar{t}$ event, we 
use $\rm \epsilon_b =0.66$,
whereas for other cases it is set to $\rm \epsilon_b =0.55$.
The total background cross-section is found to be 750~fb and 730~fb 
corresponding to two sets of selections as described 
in Tables~\ref{tab:lmevt} and \ref{tab:hmevt} respectively. 
We summarize signal significances, as defined $\rm \frac{S}{\sqrt{B}}$, where S and B are the total number of signal and background events,
corresponding to  five benchmark points in Table~\ref{tab:signi} and 
for two choices of integrated 
luminosities $\lumi=$300~$\invfb$ and 3000~$\invfb$.  
It is to be noted that in background estimation, the contribution due
to QCD is not taken into account, where jets and mis-measurement
of jets can fake as b-jets and $\rm \MET$ respectively, which is beyond the scope of this current analysis.

\begin{table}[H]
	\centering
	\caption{Signal significances for benchmark points(BP2-BP6) 
		for two luminosity options.}
	\begin{tabular}{c c c c c c}
\hline  
 & BP2 & BP3 & BP4 & BP5 & BP6 \\ 
\hline 
$\rm \frac{S}{\sqrt{B}}(\mathcal{L}=300~ fb^{-1}$) 
& 11   &  14 & 8 & 7 & 3.5  \\

$\rm \frac{S}{\sqrt{B}} (\mathcal{L}=3000~ fb^{-1}$)
& 35  &  44 & 25 & 22 & 11\\
\hline

\end{tabular}
 
	\label{tab:signi}
\end{table}
Remarkably, the significances are more than 5$\sigma$ even for lower 
luminosity option.

\begin{table}[H]
	\begin{center}
		\caption{Signal and background events for very `low mass' benchmark 
			point(BP1) in the di-lepton scenario.}
		\begin{tabular}{c c c c c c c}
 \hline   & BP1  & $\rm t\bar{t}$  & DY + jets & W+jets  & WW+jets  & WZ+jets \\ 
 \hline $\rm \sigma\times \epsilon_{BR}$~(pb)  & 1.2  & $598$  & $4242$ & 5$\times 10^4$ & $116$  & $51$  \\ 
  $\rm \MET>30 ~GeV$  & {$0.8$}  & {$371.7$}  & {$314.2$}& {$10771$} & {$46.8$}  & {$23.7$}  \\ 
  $\rm n_j\geq 1$  & {$0.74$}  & {$371.1$}  & {$301.7$}& {$10516$} & {$45.2$} & {$23.3$}  \\ 
  $\rm N(lepton)=2$ & {$0.005$}  & {$15.2$}  & {$16.5$}& {$0.2$} & {$1.1$}  & {$0.4$}  \\ 
  $\rm M_{\ell\ell}<10~ GeV$  & {$0.0032$}  & {$0.08$}  & {$0.11$}& {$0.07$} & {$0.01$}  & {$0.001$} \\ 
  b-veto  & {$0.0032$}  & {$0.024$}  & {$0.11$} &  {$0.07$} & {$0.01$}  & {$0.001$}  \\ 
  $\rm \sigma$$\times$ K-factor & {$0.006$} & {$0.034$} & {$0.14$}& {$0.1$} & {$0.02$} & {$0.002$}\\
  \hline
\end{tabular}

		\label{LM}
	\end{center}
	
\end{table}
In Table \ref{LM}, we present signal cross-section yield, presented by
Eq.~\ref{eq:lowmass}, corresponding 
to lower range of Higgs boson masses. 
Cross-sections (LO) shown in the 1st row are computed 
using MadGraph5 aMC@NLO-2.6.4(MG5NLO) subject to cut $\rm p_T^j>20$~GeV, 
whereas in the subsequent 
rows, those are presented   
after each set of selection cuts, as shown.  Notice
the severe effect of selection cut of invariant mass of lepton pair.
Finally, at the last row, we present cross-sections, multiplying 
respective K-factors to take care of higher order effects.  
Similar K-factors are used for signal process and $\rm t \bar t$ process, 
whereas for DY process it is taken to be 1.3\cite{H.:2020ecd}.
For electroweak processes, W+jets, WW+jets, WZ+jets K-factors are
considered to be 1.42~\cite{Boughezal:2015dva}, 1.8~\cite{Gehrmann:2014fva} 
and 2.07~\cite{Grazzini:2016swo} respectively.
We find the dominant background contributions are mainly due 
to the t$\rm \bar t$, DY and W+jets processes. 
We have also checked the background contribution due to 
$\Upsilon$ and $J/\psi$ production process, and found to be negligible
attributing to comparatively harder $\rm \MET$ cut. 

The total background cross-section are obtained to be $\sim$ fb and signal 
significance turns out to be, $\frac{S}{\sqrt{B}} \sim$ 6(19) for 
integrated luminosity options 300~$\invfb$
(3000~$\invfb$).

\section{Summary}

Various experiments for DM searches have excluded a substantial range of its mass. However, the DM candidate with a very low mass is still a viable option to explain the right relic density of our universe. 
In this study, we explore the scenario with a light DM candidate in the framework of the NMSSM which is constructed to address the $\mu$-problem of the MSSM by adding one additional singlet Higgs scalar with the two Higgs doublets. In this model, the lightest neutralino, assumed to be a LSP of very low
mass, is offered as a DM candidate. The significant presence of singlino component in the lightest neutralino helps to evade constraints on DM-nucleon scattering cross-section imposed by 
several experiments. In this proposed scenario, the DM annihilation takes place primarily via resonant process mediated by singlet-like light Higgs bosons, which decay to a pair of fermions in the final state.
Thus, the suppressed interaction between singlino-like neutralino and singlet-like Higgs scalars 
is responsible to overcome the stringent constraint due to observed relic abundance. 
Notably, the light non-SM-like Higgs bosons play a role as a portal between the non-SM and the 
SM sectors present in the initial and final states of the annihilation
process respectively.

A representative numerical scan of model parameters is performed taking into account
various existing experimental constraints to
identify compatible region corresponding to our proposed DM 
solution. This naive numerical study indicates that the NMSSM parameters of our interest are of the range,  
$\kappa \sim 10^{-3}-10^{-2}  $, $\rm \lambda \sim 0.1-0.3$,
$\rm |A_{\kappa}|\sim 10-100$~GeV and $\rm A_{\lambda} \gsim 800 GeV$, 
which are very close to our speculation based on analytical arguments, as discussed 
in sections \ref{sec:model} and \ref{sec:scan}. Allowed 
regions of corresponding parameters are demonstrated for the sake of illustration.    

There are various interesting phenomenological implications at the LHC of the singlino-like DM candidate, 
which are complimentary to direct searches of it in recoil experiments. For instance, in this current study, we have explored the discovery potential of such low mass DM candidate at the LHC corresponding to its high luminosity options. The DM particle is considered to be produced through SM Higgs production. 
The SM Higgs boson, produced via the standard dominant gluon-gluon fusion process, decays to a pair of  light non-SM-like Higgs bosons. Subsequently, one of the light Higgs bosons decays to a pair of DM particle resulting in missing energy, whereas the other one decays primarily to a pair of, either b quarks or $\tau$ leptons, depending on its mass. In order to make the final state more boosted, we required one extra jet
accompanied with SM Higgs boson production. The signal final state is characterized by a HJ, and missing transverse energy accompanied with at least one untagged jet. The HJ is tagged by employing sophisticated MD technique. For the lower range of lighter Higgs boson of mass $<$10 GeV, we consider its decay to a pair of $\tau$ leptons,  which eventually considered to decay in the leptonic channel leading to a final state 
with two leptons of opposite charge along with missing transverse energy and at least one untagged jet.
For the sake of presentation of signal sensitivity, six benchmark points are selected covering all possible mass ranges. Detailed simulation for both the signal and backgrounds are carried out taking into account the detector effects by using Delphes. Investigating both signal and background event characteristics, we 
have developed search strategy to suppress background contribution corresponding to a given range of light Higgs boson masses.
We found that for medium and higher combination of LSP and light Higgs boson masses, as presented by benchmark points, the sensitivity is more than $5\sigma$ for an integrated luminosity ${\cal L}$=300~$\invfb$, and for high integrated luminosity option, ${\cal L}$=3000~$\invfb$, the it further goes up. This study clearly indicates that the discovery potential for most of the mass range which are consistent with DM solution is very promising with a reasonably high luminosity option of the LHC. We have also carried out a simulation for lower mass range of non-SM Higgs boson, less than 10 GeV, of Higgs boson in leptonic final states. Our naive study shows a interesting results of achieving signal sensitivity with a
reasonable significance. It is to be noted that in this study the uncertainty due to systematics are not considered, which is beyond the scope of the present study. We conclude that the singlino-like LSP may be a very good viable candidate for DM corresponding to its lower mass range, and its signature at the LHC is also robust with a reasonably promising discovery potential for future luminosity options. However non observation of any signal event necessarily does not rule out completely the NMSSM with low mass singlino LSP scenario, instead it constrains the combination of cross-sections and related branching ratios. One requires more detailed and exhaustive scan in order to conclude about the complete exclusion of this scenario.\\

{\bf Acknowledgements}

One of the authors, A.R, wants to give special thanks to Aravind H. Vijay, Suman Chatterjee, Soham Bhattacharya and Saikat Karmakar for very useful discussions and valuable suggestions.
\section{Appendix}

\textbf{ A. DM annihilation cross-section}\\\\
\textbf{A.1 Annihilation through s-channel scalar light Higgs:}
The cross-section for the process {$\rm \tilde{\chi}_1^0\tilde{\chi}_1^0 \rightarrow H_1\rightarrow f\bar{f}$} is given by\cite{Ellis:1999mm,Nihei:2002ij}
\begin{equation}
\rm{\sigma_{f\bar{f}}^{H_1}=\frac{\omega_{f\bar{f}}^{H_1}(s)}{s^{1/2}p_1(s)}},\;\;\;\rm{\omega_{f\bar{f}}^{H_1} = \frac{g_{f\bar{f}H_1}^2\: g_{\tilde{\chi}_1^0\tilde{\chi}_1^0 H_1}^2}{(s-m_{H_1}^2)^2+m_{H_1}^2\Gamma_{H_1}^2}\frac{(s-4m_{\tilde{\chi}_1^0}^2)(s-4m_f^2)}{16\pi}\sqrt{1-\frac{4m_f^2}{s}}} ,
\end{equation}
Where,  $\rm m_{H_1}$ and $\rm \Gamma_{H_1}$ are mass and decay width of $\rm H_1$ respectively; $\rm p_1(s)=p_2(s)=\frac{1}{2}\sqrt{s-4m_{\N0_1}^2}$ is the magnitude of 3-momentum of the incoming DM particles in CM frame. $\rm g_{f\bar{f}H_1}$, $\rm g_{\tilde{\chi}_1^0\tilde{\chi}_1^0 H_1}$ are $\rm f\bar{f}H_1$ and $\rm \tilde{\chi}_1^0\tilde{\chi}_1^0 H_1$ couplings, respectively. The coupling $\rm g_{\tilde{\chi}_1^0\tilde{\chi}_1^0 H_1}$ is given in Eq.\ref{eq:ghnn}, and $\rm g_{f\bar{f}H_1}$ can be written as~\cite{Ellwanger:2009dp},
\begin{equation}
\rm  g_{t\bar{t}H_1/c\bar{c}H_1} = -\frac{m_{t/c} S_{12}}{\sqrt{2}v\sin\beta},\; 
g_{b\bar{b}H_1/\tau\tau H_1}=\frac{m_{b/\tau} S_{11}}{\sqrt{2}v\cos\beta},
\label{eq:ghff}
\end{equation}
with $\rm S_{ij}$ defined by Eq.~\ref{eq:CPevenHiggs}, $\rm m_f$ is the mass of fermion f.\\\\
\textbf{A.2 DM annihilation through s-channel pseudo-scalar light Higgs:}
Using similar notations, only replacing $\rm H_1$ by $\rm A_1$ (light pseudoscalar Higgs), we have the squared amplitude given by\cite{Ellis:1999mm,Nihei:2002ij}:
\begin{equation}
\rm{\sigma_{f\bar{f}}^{A_1}=\frac{\omega_{f\bar{f}}^{A_1}(s)}{s^{1/2}p_1(s)}},\;\;\; \rm{\omega_{f\bar{f}}^{A_1} = \frac{g_{f\bar{f}A_1}^2\: g_{\tilde{\chi}_1^0\tilde{\chi}_1^0 A_1}^2}{(s-m_{A_1}^2)^2+m_{A_1}^2\Gamma_{A_1}^2}\;\frac{s^2}{16\pi}\sqrt{1-\frac{4m_f^2}{s}}}.
\end{equation}
The coupling $\rm g_{\tilde{\chi}_1^0\tilde{\chi}_1^0 A_1}$ is given in Eq.~\ref{eq:gann}, and $\rm g_{f\bar{f}A_1}$ has similar structure as Eq.\ref{eq:ghff} except components of pseudoscalar mass matrix $\rm P_{ij}$ (Eq.~\ref{eq:CPodd}), replacing $\rm S_{ij}$.\\
Then the ``thermally averaged pair-annihilation cross-section times velocity'', $\rm \left\langle \sigma v\right\rangle $, can be obtained as\cite{Ellis:1999mm}
\begin{equation}
\rm{\left\langle \sigma v\right\rangle = \left[ \frac{1}{m_{\tilde{\chi}_1^0}^2}\left( 1-\frac{3T}{m_{\tilde{\chi}_1^0}^2}\right) \omega(s)\right]_{s\rightarrow 4m_{\tilde{\chi}_1^0}^2+6m_{\tilde{\chi}_1^0}^2T}+ O(T^2)},
\end{equation}
Where $\rm \omega(s)$ is $\rm \omega_{f\bar{f}}^{H_1}(s)$ or $\rm \omega_{f\bar{f}}^{A_1}(s)$ and T is temperature.\\\\
\textbf{B. DM-nuclei scattering cross-section}\\
WIMP-nucleon scattering cross-section is measured from the recoil of nucleon when a dark matter hits it. Generally, the velocity of these DM particles, around earth, is expected to very small ($\sim 0.001 c$). For 10 GeV DM mass, its momentum should be $\sim$10 MeV, and maximum momentum transfer is $\simeq$10 MeV. So WIMP-nucleon elastic scattering cross-section is calculated in the limit of zero momentum transfer.\\
For majorana fermion DM, the effective lagrangian can be written as\cite{Barger:2008qd,Belanger:2008sj}:
\begin{eqnarray}
\rm{\mathcal{L}= \lambda_{N}\bar{\chi}\chi \bar{\psi}\psi+i\kappa_1 \bar{\chi}\chi \bar{\psi}\gamma_5 \psi + i\kappa_2 \bar{\chi}\gamma_5\chi \bar{\psi} \psi + \kappa_3 \bar{\chi}\gamma_5\chi \bar{\psi}\gamma_5\psi+ \kappa_4 \bar{\chi}\gamma_{\mu}\gamma_5\chi \bar{\psi}\gamma_{\mu}\psi} \nonumber\\
\rm +\zeta_{N}\bar{\chi}\gamma_{\mu}\gamma_5\chi \bar{\psi}\gamma^{\mu}\gamma_5\psi
\end{eqnarray}
It can be shown that, in the zero momentum transfer limit, $\rm \bar{u}\gamma_5 u$ (u can be $\rm \chi$ or $\rm \psi$) vanishes  and also the time component of $\rm \bar{u}\gamma_{\mu}\gamma_5 u$ and the space component of $\rm \bar{u}\gamma_{\mu} u$ tends to zero.\\\\
\textbf{B.1 SI cross-section}\\
So the effective spin independent interaction can be written as :
\begin{equation}
\rm{\mathcal{L}^{SI}= \lambda_{N}\bar{\chi}\chi \bar{\psi}\psi}
\end{equation}

In our case, this spin-independent scattering cross-section ($\rm \sigma_{SI}$) of $\rm \tilde{\chi}_1^0$ with nuclei dominantly happens through exchange of scalar Higgs bosons. When $\rm \tilde{\chi}_1^0$ is singlino-like, we can write the scattering cross-section approximately as \cite{Gunion:2005rw}
\begin{equation}
\rm{\sigma_{SI} \simeq \frac{1}{\pi m_{H_1}^4}\left(\frac{m_p m_{\tilde{\chi}_1^0}}{m_p + m_{\tilde{\chi}_1^0}}\right)^2 g_{\tilde{\chi}_1^0\tilde{\chi}_1^0 H_1}^2 \left(\sum_{q=d,s,b}^{}\frac{m_q S_{11}}{cos\beta}\left\langle N\left| q\bar{q}\right|N \right\rangle +\sum_{q=u,c}^{}\frac{m_q S_{12}}{sin\beta}\left\langle N\left| q\bar{q}\right|N \right\rangle \right)^2 }
\end{equation}
Where, $\rm \left\langle N\left| q\bar{q}\right|N \right\rangle$ are the matrix element over the atomic nuclear states, $\rm m_p$ is the mass of the nuclei. Other notations have usual meanings.\\\\
\textbf{B.2 SD cross-section}\\
The effective lagrangian in this case can be written as:
\begin{equation}
\rm{\zeta_{N}\bar{\chi}\gamma_{\mu}\gamma_5\chi \bar{\psi}\gamma^{\mu}\gamma_5\psi}
\end{equation}
Here DM-nucleon scattering can be mediated in t-channel by Z-boson or squark mediator (I denote it as V, with mass $\rm m_V$) . The cross-section in this case is becomes :
\begin{equation}
\rm{\sigma_{SD} \simeq \frac{4}{\pi m_V^4}\left(\frac{m_p m_{\tilde{\chi}_1^0}}{m_p + m_{\tilde{\chi}_1^0}}\right)^2\left(\frac{J_A+1}{J_A}\right) g_{\tilde{\chi}_1^0\tilde{\chi}_1^0 V}^2 \left(\sum_{q=u,d,c,s,b}^{}\left\langle N\left| q\bar{q}\right|N \right\rangle \right)^2 \left(\zeta_p S_p^A+\zeta_nS_n^A \right)^2}
\end{equation}
Where $\rm J_A$ is the angular momentum of the nucleus with A nucleons and $\rm S_N^A$ are the expectation value of the spin content of nucleon type N (n or p).\\

\bibliographystyle{utphys.bst}
\bibliography{paperv2.bib}


\end{document}

\end{document}